\documentclass[11pt]{article} 

\usepackage{amsmath,amsthm,latexsym,amssymb,amsfonts,epsfig}

\usepackage{cite}
\usepackage{hyperref}
\hypersetup{
    colorlinks,%
    citecolor=blue,%
    filecolor=blue,%
    linkcolor=blue,%
    urlcolor=blue
}

\newtheorem*{Definition}{Definition}
\newtheorem*{Lemma}{Lemma}

\newcommand{\be}{\begin{equation}}
\newcommand{\ee}{\end{equation}}
\newcommand{\ba}{\begin{eqnarray}}
\newcommand{\ea}{\end{eqnarray}}

\title{{\sf Hamiltonian Renormalisation I:}\\
{\sf Derivation from Osterwalder-Schrader Reconstruction}} 
\author{
{\sf T. Lang}$^1$\thanks{{\sf 
thorsten.lang@gravity.fau.de}},
{\sf K. Liegener}$^1$\thanks{{\sf 
klaus.liegener@gravity.fau.de}},
{\sf T. Thiemann}$^1$\thanks{{\sf 
thomas.thiemann@gravity.fau.de}}\\
\\
{\sf $^1$ Inst. for Quantum Gravity, FAU Erlangen -- N\"urnberg,}\\
{\sf Staudtstr. 7, 91058 Erlangen, Germany}\\
}
\date{{\small\sf \today}}

\makeatletter
\@addtoreset{equation}{section}
\makeatother

\begin{document} 

\maketitle

{\sf
\begin{abstract}
A possible avenue towards a non-perturbative Quantum Field Theory (QFT) on 
Minkowski space is the constructive approach which employs 
the Euclidian path integral formulation, in the presence of both 
ultraviolet (UV) and infrared (IR) regulators, as starting point. The
UV regulator is to be taken away by renormalisation group techniques 
which in case of success leads to a measure on the space of generalised
Euclidian fields in finite volume. 
The removal of the IR regulator corresponds to  
the thermodynamic limit of the system in the statistical physics sense.
If the resulting measure obeys the Osterwalder-Schrader axioms,
the actual QFT on Minkowski space is then obtained by Osterwalder-Schrader
reconstruction.

In this work we study the question whether it is possible to reformulate 
the renormalisation group non-perturbatively directly at the operator 
(Hamiltonian) level. 
Hamiltonian renormalisation would be the natural route to follow if 
one had easier access to an interacting Hamiltonian operator rather than 
to a path integral, at least in the presence of UV and/or IR cut-off, 
which is generically the case in complicated gauge theories such as 
General Relativity.

Our guiding principle for the definition of the direct Hamiltonian 
renormalisation group is that it results in the same continuum theory
as the covariant (path integral) renormalisation group. This makes the 
present work different from other approaches to Hamiltonian renormalisation.
In order to achieve this, we invert the Osterwalder-Schrader reconstruction,
which may be called Osterwalder-Schrader {\it construction} of a 
Wiener measure from the underlying Hamiltonian theory. The resulting 
correspondence between reflection positive measures and Osterwalder-Schrader
data consisting of a Hilbert space, a Hamiltonian and a ground state vector
allows us to monitor the effect of the renormalisation flow of measures
in terms of their Osterwalder-Schrader data which motivates a natural
direct Hamiltonian renormalisation scheme. 
\end{abstract}

\newpage

\tableofcontents

\newpage

~\\
{\bf Notation:}\\
\\
In this paper we will deal with quantum fields in the presence of an 
infrared cut-off $R$ and with smearing functions of finite time support 
in $[-T,T]$. The spatial ultraviolet cut-off is denoted by $M$ and has  
the interpretation of the number of lattice vertices in each spatial 
direction. We will mostly not be interested in an analogous temporal 
ultraviolet cut-off $N$ but sometimes refer to it for comparison 
with other approaches. These quantities allow us to define dimensionful
cut-offs $\epsilon_{RM}=\frac{R}{M},\; \delta_{TN}=\frac{T}{N}$. In 
Fourier space we define analogously 
$k_R=\frac{2\pi}{R},\; k_M=\frac{2\pi}{M},\;   
k_T=\frac{2\pi}{T},\; k_N=\frac{2\pi}{N}$.

We will deal with both instantaneous fields, smearing functions and Weyl 
elements as well as corresponding temporally dependent objects. 
The instantaneous objects are denoted by lower case letters 
$\phi_{RM}, \; f_{RM},\; w_{RM}[f_{RM}]$, the temporally dependent ones by
upper case ones
$\Phi_{RM}, \; F_{RM},\; W_{RM}[f_{RM}]$. As we will see, smearing functions
$F_{RM}$ with 
compact and discrete (sharp) time support will play a more fundamental role 
for our purposes than those with a smoother dependence. 

Osterwalder-Schrader reconstruction concerns the interplay between 
time translation invariant, time reflection invariant and reflection positive 
measures (OS measures)
$\mu_{RM}$ on the space of history fields $\Phi_{RM}$ and their corresponding
Osterwalder-Schrader (OS) data ${\cal H}_{RM}, \Omega_{RM}, H_{RM}$ where   
${\cal H}_{RM}$ is a Hilbert space with cyclic (vacuum) vector $\Omega_{RM}$ 
annihilated by a self-adjoint Hamiltonian $H_{RM}$. Together, the vector
$\Omega_{RM}$ and the scalar product $\langle.,.\rangle_{{\cal H}_{RM}}$ define 
a measure $\nu_{RM}$ on the space of instantaneous fields $\phi_{RM}$. 

Renormalisation consists in defining a flow or sequence 
$n\to \mu^{(n)}_{RM},\; n\in \mathbb{N}_0$ for all
$M$ of families of measures $\{\mu^{(n)}_{RM}\}_{M\in \mathbb{N}}$. 
The flow will be defined in terms of a coarse graining or embedding 
map $I_{RM\to M'},\; M<M'$ acting on the smearing functions and satisfying 
certain properties that will grant that 1. the resulting fixed point family 
of measures, if it exists, is cylindrically consistent and 2. the flow 
stays within the class of OS measures. Fixed point quantities are denoted 
by an upper case $^\ast$, e.g. $\mu^\ast_{RM}$.

\newpage

\section{Introduction}
\label{s1}

Renormalisation in Quantum Field Theory (QFT) comes in two flavours, namely
perturbative and non-perturbative schemes. In the perturbative formulation
it is a matter of taste whether one uses operator or path integral methods
as both are based on the underlying free theory for which there is 
a precise bijection between both methods. Roughly speaking, the path 
integral of a free theory can be obtained by Trotter product techniques 
from the operator formalism (Feynman-Kac formula). Conversely, the operator 
formalism is recovered from the path integral of the free theory by 
Osterwalder-Schrader reconstruction.

On the other hand, non-perturbative 
approaches, at least in 4 dimensions, are almost entirely based on the 
path integral approach, in particular in its Euclidian formulation following
the constructive QFT programme \cite{GJ87, Fr78, Riv00, Sim74}. 
The reason for this is that in the presence 
of both ultraviolet (UV) and infrared (IR) cut-off, it is very easy to 
postulate a rigorously defined path integral for the interacting theory
as the system retains only a finite number of degrees of freedom and 
at least for QFT on Minkowski space a path integral defined by the Lebesgue 
measure on the space of those finitely many degrees of freedom times 
the exponential of minus the Euclidian action is a well motivated 
starting point. The challenge is then to show that one can take first 
the UV and then the IR 
regulator away, resulting in a rigorously defined Euclidian measure 
obeying the Osterwalder-Schrader axioms, most prominently reflection 
positivity \cite{OS72,OSei78,simon73}. The IR cut-off removal corresponds to the thermodynamic
limit in the statistical physics sense which is what Euclidian QFT 
is and is usually less problematic than the UV cut-off removal.

The technical tool for the UV cut-off removal is the renormalisation
programme, which we will review, in the Euclidian setting, 
for a rather general theory in a subsequent section. In our terminology,
a QFT with a short distance or finite resolution
cut-off is simply a one parameter family
of more or less motivated probability 
measures $\epsilon\mapsto \mu^{(0)}_\epsilon$ 
that typically arise by ad hoc discretisations on finite lattices of 
finite resolution $\epsilon$. Here $\epsilon$ 
may be defined via a background metric or not, in the presence of an IR cut-off
the spacetime available to the quantum field is compact and hence $\epsilon$
can be defined background independently by counting the number of degrees
of freedom. It is important to note, that the ad
hoc prescription for building $\mu^{(0)}_\epsilon$, while generically
plagued by ambiguities (``couplings'' on which the measure depends 
parametrically) is the same for all 
$\epsilon$. The measure $\mu^{(0)}_\epsilon$ is supposed to produce 
all Schwinger functions from the generating functional 
$\mu^{(0)}_\epsilon(\exp(i\Phi_\epsilon[F_\epsilon]))$ by (functional) 
derivation at $F_\epsilon=0$  with respect to $F_\epsilon$. Here 
$\Phi_\epsilon$ is the discretised field at cut-off $\epsilon$ over which 
$\mu^{(0)}_\epsilon$ integrates and $F'_\epsilon$ is the restriction
to the lattice vertices of a continuum 
smearing function $F_\epsilon$ which however only resolves scales at or 
above $\epsilon$. In terms of the undiscretised or continuum field $\Phi$
one has the relation $\Phi[F_\epsilon]=:\Phi_\epsilon[F'_\epsilon]$
between the smeared fields. As we will see, the details of the relation
between $F_\epsilon$ and $F'_\epsilon$ is what defines a block spin 
or better coarse graining transformation. If one had a continuum measure $\mu$ 
integrating over continuum fields 
$\Phi$, then one would identify the generating functional at $\epsilon$ with
$\mu(\exp(i\Phi(F_\epsilon))$. In other words, the $\mu^{(0)}_\epsilon$ 
would define the cylindrical projections of $\mu$. However, for this to be the 
case, the family $\mu^{(0)}_\epsilon$ must be cylindrically consistent:
We can view $F_\epsilon$ as a special case of a function 
$\tilde{F}_{\epsilon'}$ at a scale 
$\epsilon<\epsilon'$, that is, $F_\epsilon=\tilde{F}_{\epsilon'}$ as functions
in the continuum. They will give rise to finite resolution functions 
$F'_\epsilon,\; \tilde{F}'_{\epsilon'}$. Now by construction the 
continuum measure will assign to both functions the same generating 
functional but that is not automatically the case for the generating
functionals in terms of the family $\mu^{(0)}_\epsilon$. 

Renormalisation now consists in constructing
a sequence $\mu^{(n)}_\epsilon$ of measure families where one 
obtains $\mu^{(n+1)}_\epsilon$ from $\mu^{(n)}_{\epsilon/2}$ by 
integrating out the degrees of freedom at scale $\epsilon/2$ that 
do not contribute to scale $\epsilon$. If the sequence converges or 
at least has a fixed point (accumulation point) family $\mu^\ast_\epsilon$ 
then by construction that family is consistent. Then the usual 
extension theorems \cite{Yam75} can be consulted in order to show that the family 
extends to a continuum measure $\mu^\ast$. One then must show 
that $\mu^\ast$ obeys the Osterwalder-Schrader axioms. To be a viable 
physical theory, all but a finite number of the ambiguous parameters (the 
so-called relevant ones),
that parametrise the original family $\mu^{(0)}_\epsilon$, must have been
fixed pointed during the renormalisation process. All of this will be made
precise during the course of the paper.

How does the Hamiltonian description fit into this picture? We restrict
the block spin transformations defining the renormalisation flow by the 
requirement 
that the measures $\mu^{(n)}_\epsilon$ for each $n\in \mathbb{N}_0, \epsilon>0$
satisfy at least a subset of the Osterwalder-Schrader axioms sufficient 
in order to reconstruct quantum theory (see e.g. \cite{AMMT99} for such a subset).
This will impose a certain restriction on the choice of the initial family
$\mu^{(0)}_\epsilon$ as well as the choice of coarse graining map. 
Under these circumstances we obtain for each $n,\epsilon$ a triple 
$({\cal H}^{(n)}_\epsilon,\; H^{(n)}_\epsilon,\; 
\Omega^{(n)}_\epsilon)$ consisting of a Hilbert space, a self-adjoint 
and positive semi-definite 
Hamiltonian operator thereon and a cyclic vacuum vector which is a zero
eigenvector for the Hamiltonian. We will call them Osterwalder-Schrader
(OS) data in what follows. Therefore, from the measure theoretic 
perspective, Hamiltonian renormalisation consists in precisely 
the flow of this family
of triples. Its  fixed point $({\cal H}^\ast,\; H^\ast,\;\Omega^\ast)$
are the Osterwalder-Schrader data for $\mu^\ast$ 
provided that the measure family has a fix point $\mu^\ast$ satisfying 
the Osterwalder-Schrader (OS) axioms.

In this paper we ask the natural question, whether the renormalisation 
sequence of OS data really has to be constructed via the measure theoretic
framework or whether there is a more direct route that remains entirely
within the Hamiltonian setting. In fact, the initial 
measure family $\mu^{(0)}_\epsilon$ can only be ``guessed'' for a tiny
subset of theories that are of the form: Lebesgue measure times exponential 
of minus the negative of the classical Euclidian action. This is the 
subset of theories for which the classical Hamiltonian is 0. not constrained 
to vanish, 1. a quadratic 
polynomial in the bosonic canonical momenta with coefficients that are 2. 
independent 
of the configuration field and 3. independent of time. If the zeroth 
assumption fails to hold, then we consider a generally covariant theory
that is spacetime diffeomorphism invariant and the measure will
typically receive very complicated corrections from the gauge fixing
Dirac-bracket determinant (Fadeev-Popov determinant), see e.g. \cite{HT94,Vyt94,MRAV}. 
If the  
first condition is not met, then the momenta cannot even be integrated out 
and one must use a path integral formulation which is based on a measure 
on a history phase space rather than history configuration space 
\cite{HT94}. If the second condition is violated, one can integrate out the
momenta, but the Lebesgue measure receives a non-trivial Jacobian. Finally,
if the third assumption is false, then a Wick rotation to imaginary 
time is generically impossible. This is the case of QFT in curved spacetime 
since general background spacetime metrics  
do not admit a gauge (coordinate system) in which the metric is analytic
in the coordinates.
In this paper we will deal with theories that underlie restriction 1. and 
3. but not necessarily the zeroth and second. This includes quantum gravity 
coupled to standard matter when suitable gauge fixing of spacetime 
diffeomorphism invariance 
employing matter reference systems is in place \cite{5}. This works, 
because in that case 
the metric is a dynamical field with no explicit time dependence. 
More generally then, we restrict attention to conservative systems (which
is no actual restriction, every system becomes conservative when including the 
background as a dynamical field) where possible diffeomorphism invariance 
has been gauge fixed.

Given such a set-up,
the construction of an initial family of triples consisting
of Hilbert space, Hamiltonian and vacuum is relatively straightforward in 
this setting. If we are interested in the continuum 
OS data, then the route via the construction of the initial measure family 
and its renormalisation looks like a very indirect and complicated
approach that one would like to circumvent.
In order to perform Hamiltonian renormalisation directly without using 
the OS measure, it is necessary to understand how the bijection between
the OS measure and the OS data actually works. In \cite{AMMT99}, the part 
of the proof, 
usually given for Minkowski space QFT, that suffices to reconstruct 
the OS data from the measure, has been adapted to a more general setting 
including 
generally covariant QFT. In this paper we will develop the opposite direction
and show that both maps are inverses of each other when the theory has been
suitably formulated. After having performed 
this step we can try to define Hamiltonian renormalisation directly
without recourse to the functional integral in such a way that both 
renormalisation procedures are mapped into each other. We will be only 
partially successful in doing so: While we can show that for suitable 
coarse graining maps the path integral flow stays in the class of 
reflection positive measures, the construction of these measures cannot 
be entirely avoided. On the other hand, the framework suggests a 
natural modification of the path integral induced flow giving rise to
a direct flow of the OS data without using path integrals which we will
demonstrate to have the same fixed points in the case of free field theories. 
In that sense 
we differ from earlier, related suggestions in the literature (see
e.g. \cite{Bal89a, Bal89b,Dim12a,Dim12b,Dim12c,Dim12d,WK73,BW74,Wil75,book:971208,book:1081264,T-1984,article,K-1980,book:970374,Davi-1979,book:16786,H-1976} or 
for applications to quantum gravity \cite{CZ97, MOWZ06,BD09,BDS11,GL13,DMS14,EV15,BS17} 
and references therein). Rather, our approach is closer in spirit to a purely 
Hamiltonian scheme similar to the frameworks developed by Glazek and Wilson \cite{GlaWil}
building on Wegner \cite{Wegner} for which there are vast applications in the literature
(see e.g. \cite{AlexanianMoreno, Ciobanu} and references 
therein for more recent applications within perturbative interacting QFT and quantum 
mechanics). Our considerations differ from these latter works not only 
in that the relation to the path integral 
renormalisation scheme remains more transparent. Specifically, our renormalisation
group flow will be in terms of families of 
isometric injections from Hilbert spaces of lower 
to those of higher resolution rather than block diagonalising 
unitaries with respect to higher and lower energy modes. 
Moreover, our formulation is intrinsically background independent and 
non perturbative.   

Roughly speaking, direct Hamiltonian renormalisation is parametrised by a 
family
of spatial coarse graining maps on the space of test functions. These in turn 
define 
embeddings of coarse grained Hilbert spaces. The flow of the OS data 
is such that these embeddings are isometric,  in particular 
when the fixed point is 
reached. The continuum Hilbert space is then the corresponding inductive 
limit, the vacua are mapped into each other by those isometries. 
The flow of the Hamiltonians is such that they annihilate the corresponding
flow of vacua, however, the continuum Hamiltonian defined by the 
fixed point family is {\it not} the inductive 
limit of the coarse grained Hamiltonians. Indeed, for an 
inductive family of Hamiltonian operators it is necessary that the
family of isometric injections is  equivariant with respect to the family
of Hamiltonians, however, we obtain only a weaker version of that.

This is because the fixed point family only defines a consistently 
defined quadratic form (if certain universality properties with respect to the 
chosen coarse graining map are satisfied). Such a quadratic form does not 
necessarily define an operator. By contrast,
the inductive limit construction grants the existence of such a 
Hamiltonian but would require that the Hamiltonians
preserve the image of the injections which again would be counter intuitive.
This also means that we are not in conflict with Haag's theorem which 
for an infinite number of degrees of freedom predicts difficulties when 
trying to define different Hamiltonians on the same Hilbert space. 
On the other hand, the fact 
that an interacting vacuum vector is constructed in the sense 
that the Hamiltonian quadratic form has vanishing matrix element with any 
other state increases the chance that the quadratic form in fact extends 
to an operator. Moreover, if the initial
operator family is positive and symmetric, then the flow preserves 
this property and thus results in a positive, symmetric quadratic form. 
This quadratic form then, provided it is closable, has at least the Friedrichs
extension as a self-adjoint operator. Of course, it is then a physics 
question if that self-adjoint extension is viable.    

We also note that in order to construct the Hamiltonian from the OS measure
we need that time is continuous as the Hamiltonian generates the corresponding
OS contraction semi-group. Hence, the measure renormalisation that we consider
here is one where time is always continuous. This asymmetric treatment 
of time and space is mandatory if we want to construct the Hamiltonian. \\
\\
The architecture of this paper is as follows:\\
\\

In section \ref{s2} we derive the bijection between OS measure and OS
data.

In section \ref{s3} we formulate measure renormalisation in a language that 
slightly differs from the usual one and that is geared towards Hamiltonian 
renormalisation in our sense. We present the dictionary between the usual 
language and our own.

In section \ref{s4} we derive path integral induced 
Hamiltonian renormalisation from first 
principles, that is, using the language of sections \ref{s2} and \ref{s3}.
The main input is the coarse graining map. We motivate the choice of 
physically acceptable maps starting from the classical theory where the 
role of the corresponding isometric injections of the quantum theory 
is played by classical partial symplectomorphisms. This is 
in accordance with Kijowski's programme of formulating large quantum theories 
as consistent families of small ones \cite{KO16, LaneryT14,Lanery,MO16}. This also makes it plausible 
that two choices of coarse graining maps result in unitarily equivalent 
QFT's. The Hamiltonians that one obtains via OS reconstruction from   
the path integral flow have the counter intuitive property to commute 
with the projections onto the coarse grained Hilbert spaces. We uncover the 
semigroup law of the OS contraction semigroups generated by these Hamiltonians 
as the reason responsible for that feature. This observation lies at the 
heart of defining a closely related direct Hamiltonian flow which does 
not have that counter intuitive feature.    

In section 5 we summarise our findings and discuss future applications of 
the framework developed in the present paper, in particular for 
theories that are generally covariant and thus background independent.

In appendix \ref{sb} we recall elementary notions from the theory of inductive
limits in our language for the benefit of the reader.

In appendix \ref{sc} we supply some background material on criteria when 
a Hamiltonian contraction semigroup defines a Wiener measure.
\\
This paper is the first in a series of four papers. The three companion 
papers test the framework of the present paper in a setting where we 
have full control on what the answer to various questions must be: Free 
scalar field theories on Minkowski space in arbitrary dimensions.
Note, however, that the framework 
developed here is background independent. Instead of using resolution scales 
with respect to a background metric that are used in these companion papers
one can abstract to partially ordered
and directed label sets which in fact motivated this article as we have 
applications in Loop Quantum Gravity in mind \cite{Ash91, Rov04,Thi07,GS13}.
The language developed here could also be applied directly to the 
asymptotic safety approach to quantum gravity \cite{ReuSau07,Perc10,ReuSau12,Eich17}, a task 
to which we want to come back in the future. 
 
In the first companion paper \cite{21a} we analytically derive the Hamiltonian 
renormalisation flow for the 1+1-dimensional, massive Klein Gordon model and 
compute its fixed points. The restriction to 1+1 dimensions helps illustrating
the basic mechanisms without exceedingly cluttering the formulae. We compare 
the direct and path integral induced flow and find a surprise, as could 
be anticipated from the general framework of the present paper: The 
path integral induced and direct Hamiltonian flow drastically differ from 
each other and only the coarse grained versions of the direct Hamiltonian
flow qualify as the matrix elements of the continuum Hamiltonian between 
coarse grained vector states. 

In the second companion paper \cite{21} we investigate further properties of 
the renormalisation flow of the 1+1-dimensional model such as i. whether 
the flow in fact converges to the fixed point computed, ii. the decay 
properties of the correlations of the fixed point Laplacian with respect to 
distance between lattice points and iii. universality properties of the 
flow (dependence on the coarse graining map and initial data).

Finally, in the third companion paper \cite{22} we consider D+1 dimensional 
free scalar quantum fields. It turns out that the analytical methods used for 
the 1+1 case can be directly transferred due to a certain factorisation 
property. The flow has a fixed point which agrees with the Euclidian 
(or rotationally) invariant
continuum theory. This enables us to study the question whether there is  
a criterion that enables one to decide at finite resolution, i.e. without 
taking the continuum limit, i.e. on a finite lattice, 
whether the flow converges to a 
Euclidian (or rotationally) invariant theory. We define such a criterion 
using irrational Euler angle techniques.

\section{Bijection between OS measure and OS data}
\label{s2}

We begin in the first subsection 
with some preliminaries that explain the class of theories considered and prepare the 
language and assumptions necessary in order that a suitable bijection between 
path integral and Hamiltonian formulation can be stated. 
In the next subsection of this section we construct the 
OS measure from OS data based on the idea of the Wiener measure or the Feynman-Kac 
formula, see e.g. \cite{Dim12d} and references therein. Then we perform the more standard 
converse procedure which we include here only for the sake of completeness 
and in order to introduce a uniform notation, see e.g. \cite{GJ87} and references therein.
In the third section we show, under the made restrictions, that both constructions are 
inverses of each other. We believe that our presentation of partly well known material
will be useful to a mixed audience, in particular, because we employ a background 
independent and generally theory independent language.

\subsection{Preliminaries}
\label{s2.0}

Let us first explain the general setting:\\
\\
We consider a general theory based on a classical Lagrangian
(or Hamiltonian) which is not explicitly coordinate dependent. We leave the 
field content open, but assume that diffeomorphism type gauge symmetries have 
been taken care of by suitable gauge fixings. In particular, even in generally 
covariant theories, after suitable gauge fixing of the timelike diffeomorphisms, 
there is a preferred notion of time, see e.g. \cite{KT91,BK95,HP11,GT12} and references therein. 

In more detail, as we are interested in 
situations with a well posed classical initial value formulation,
the underlying spacetimes under consideration are globally hyperbolic
so that points in the spacetime manifold $M$ can be denoted by $(t,x)$ where 
time $t$ takes real values and $x$ is a point in some fixed submanifold 
$\sigma$ of
$M$ of codimension one. It follows from our assumptions that all time zero 
fields, collectively denoted by $\phi(x)$ and their canonical momenta, 
collectively denoted by $\pi(x)$, are defined with respect to a physical
coordinate system. We further assume that $(\phi,\pi)$ have standard
canonical brackets which is not easy to achieve due to the gauge fixing 
conditions which typically add corrections to the canonical (Dirac) bracket, but
is possible for suitable reference matter fields, see e.g. \cite{GT12}.

On the other hand, usual Yang-Mills type gauge symmetries need not have
to be gauge fixed. Instead, we assume that in this case one employs 
gauge covariant versions of the usual Weyl elements \cite{Thir81,BR04}
$w(f,g):=\exp(i[\phi(f)+\pi(g)])$ that one uses in scalar field theories.
For Bose fields, $(\phi,\pi)$ and $(g,f)$ are 
geometrically both a  
pair consisting of a spatial scalar density of weight 
zero and one respectively such that the pairing 
$\phi(f)=\int_\sigma \; d^Dx\; f\; \phi$ is invariant under active spatial
diffeomorphisms ($f,F$ are real valued). 
Thus, for instance for a non-Abelian Yang-Mills theory
or General Relativity when formulated in terms of connection variables 
\cite{AB-variables1,AB-variables2,AB-variables3} one will utilise Wilson loop 
type holonomy variables for the 
spatial connection $\phi$ and exponentiated non-Abelian electric flux variables
for the conjugate electric field $\pi$. We will treat the case of Fermi 
fields in a future publication.

In order to be theory independent, we will talk about the corresponding 
generalised Weyl elements $w(f,g), \; w(f):=w(f,g=0)$ for the canonical 
formulation 
in what follows as well as the canonical Hamiltonian $H=H(\phi,\pi)$ 
which is supposed to be a positive functional of $\phi,\pi$ without explicit 
coordinate dependence, i.e. the system is conservative. 

We are interested in (non-degenerate) representations of the\footnote{Whether 
or not $\mathfrak{A}$ can be equipped with a unique $C^\ast-$algebra 
structure on purely algebraic grounds such as in the case of 
symplectic vector spaces \cite{BR04} will not be investigated here, see 
e.g. \cite{Fleisch99,Fleisch06} for the case of Yang-Mills fields.} $\ast-$Algebra
$\mathfrak{A}$ generated by the Weyl elements 
that supports the Hamiltonian $H$ as a self-adjoint operator. In the case of 
an interacting quantum field theory this will generically require ultra-violet and 
infra-red regularisations. With such regulators in place, let
$t\mapsto \alpha_t$ denote the one parameter group of (outer) automorphisms 
of $\mathfrak{A}$ corresponding to the classical 
Hamiltonian flow of $H$. As every (automatically non-degenerate 
as $\mathfrak{A}$ is unital) representation is a direct sum 
of cyclic ones \cite{BR04} and every such representation $(\rho,\;
{\cal H},\;\Omega)$ is generated by a 
positive linear functional $\omega$ by the GNS construction \cite{Haag92}
we are interested in invariant states $\omega\circ \alpha_t=\omega$.  
Namely, in this situation we have a representation by inner automorphisms  
\be \label{2.1}
\rho\circ \alpha_t=U(t)\circ \rho \circ U(t)^{-1},\;\;U(t)=:\exp(-iHt)
\ee
where the one parameter group $t\mapsto U(t)$ of unitarities is generated 
by $H$ according to Stone's theorem if $t\mapsto U(t)$ is strongly continuous,
as we will assume. Note that $\Omega$ is a zero eigenvector of $H$, i.e. 
a vacuum vector. For convenience we will assume that $H$ 
is bounded from below (without loss of generality by zero) just as the 
classical object. 

As already mentioned we will treat Fermi fields separately in a future publication
and focus on Bose fields for the rest of this work.
For bosonic theories of interest (the Hamiltonian is quadratic in the momenta), 
$\rho(w[f,t]):=U(t)\rho(w[f])U(t)^{-1}=
\rho(\alpha_t(w[f]))$ contains sufficient 
information about $\pi(g)$ as well
(take the time derivative at $t=0$) so that one can focus attention on 
formulating a quantum field theory in terms of the Heisenberg field operators 
$\rho(w[f,t])$ in what 
follows. Now it is in general expected that for generic 
representations $\rho$ of an
interacting quantum field theory the sharp time objects $\rho(w[f,t])$ 
are no longer represented as operators but rather as operator valued distributions (with respect
to the time variable) when we take the regulators away (and renormalise the fields) 
so that $\rho(w[f,t])$ necessarily also must be smeared in the time 
coordinate in order to obtain a well defined operator. 
For instance, this may be viewed as a consequence of Haag's theorem \cite{Haag92} for 
quantum fields on Minkowski space in the context of the Wightman axiomatic framework. 
Hence, the formulation of a quantum field theory based on the $\rho(w[f,t])$ of which we 
make extensive use throughout this series of papers appears to be of little relevance for 
the most interesting models, even if the triple of Osterwalder-Schrader data 
$({\cal H}, \Omega, H)$ survives the regulator limit as we assume throughout this series.

However, while this is generically 
certainly true for the $\rho(w[f,t])$ as defined starting from the 
classical time zero field $\phi$, under one rather mild technical assumption there 
exists a reformulation in terms of which sharp time field operators do make sense, albeit
for generically different fields.  
Namely, we will assume throughout the whole 
paper, that the final Hilbert space $\cal H$ that one ends up with
admits an Abelian 
$C^\ast-$subalgebra $\mathfrak{B}$ of the $C^\star-$algebra ${\cal B}({\cal H})$
of bounded operators on $\cal H$ for which the final vacuum vector state $\Omega$ that one 
ends up with is cyclic. This assumption is not very restrictive, it is easy to see 
that it holds for instance when $\cal H$ is separable but also in many non-separable 
cases. In fact, in our renormalisation flow application, the Hilbert spaces that we 
encounter at finite resolution are all separable, hence our reformulation that we are 
about to sketch certainly is valid in the applications that we have in mind in this series 
of works.

Let $\Delta(\mathfrak{B})$ be the Gel'fand spectrum of $\mathfrak{B}$ \cite{RS80}. Then 
we will call the points $\phi'$ of $\Delta(\mathfrak{B})$ {\it quantum fields at sharp time
zero}. Of course, in general the relation between the quantum sharp time zero field 
$\phi'$ and the classical sharp time zero field $\phi$ will be rather involved and 
depends on the choice of $\mathfrak{B}$. Taking into account of what was just said, we 
expect that $\phi'$ is related to a certain time average of $\phi$. Indeed, in the course 
of this series of papers we will see an instance of this effect when we renormalise 
the free Klein-Gordon field.   

Using the Gel'fand isomorphism \cite{RS80} we may think of $\mathfrak{B}$ as 
$C(\Delta(\mathfrak{B}))$, the continuous functions on the spectrum. Since the 
spectrum carries a compact Hausdorff topology, 
the positive linear and normalised 
expectation value functional $\langle\Omega,.\Omega\rangle_{{\cal H}}$ on $\mathfrak{B}$
gives rise to a probability measure $\nu$ on $\Delta(\mathfrak{B})$ by the 
Riesz-Markov theorem \cite{Rud87}. It follows that 
${\cal H}\cong L_2(\Delta(\mathfrak{B}),d\nu)$. If necessary, we can perform 
a unitary transformation  
such that the vacuum becomes the constant vector $\Omega'=1$. 
This is possible if $\Omega$ is not vanishing $\nu$ almost everywhere (a.e.). 
The new representation is given by $\rho'(.)=\Omega^{-1}\rho(.)\Omega,\;
\psi'=\Omega^{-1} \psi,\;\nu'=|\Omega|^2 \nu$ 
where we understand $\Omega$ as a multiplication operator. 

We consider a new generating set of $\mathfrak{B}$ consisting of bounded functions $w'[f']$ 
on $\Delta(\mathfrak{B})$
to which we refer as new Abelian Weyl elements. Here $f'$ belongs to a certain label set $L'$
of smearing functions on $\sigma$
(for instance functions of rapid decrease for free scalar fields) and each 
$b\in \mathfrak{B}$ is supposed to be in the finite linear span of the $\rho(w'[f'])$. 
For this 
to be consistent we require Weyl type relations among the $w'[f']$ which allow us to 
express 
$(w'[f'])^\ast$ and $w'[f_1']\cdot w'[f_2']$ as a finite linear combination of the 
$w'[f']$ again. Specifically we could have relations of the form 
 \be \label{2.19}
(w'[f'])^\ast=:w'[(f')^\ast],\;w'[f'_1] w'[f'_2]=:\sum_{\alpha=1}^N z_\alpha 
w'[f'_\alpha(f'_1,f'_2)]
\ee
where $N<\infty,\;z_\alpha\in \mathbb{C}$ are independent of $f'_1,f'_2$ and the new 
smearing function $f'_\alpha$ also is built from $f_1',f_2'$ in the same 
way for any $f'_1,f'_2$. Also, $(f')^\ast$ is not the complex conjugate of 
$f'$ (which is typically real valued) but rather determined 
by the fact that $(w'[f'])^\ast$ is the adjoint of $w'[f']$ and 
coincides with the complex conjugate $\overline{w'[f']}$ in 
our reformulated representation in which $w'[f']$ acts by multiplication. 
For a free scalar field 
of course $(f')^\ast=-f',\; N=1,\; z_1=1,\; f'_{\alpha=1}(f'_1,f'_2)=f'_1+f'_2$. 
For non-Abelian gauge theories formulated in terms of Wilson loop variables,
the relations (\ref{2.19}) are known as Mandelstam identities \cite{OSei78}.  

The Weyl elements at sharp time $w'[f',t]=U(t) \rho(w'[f']) U(t)^{-1}$ as a product 
of three bounded operators now make perfect sense 
as operators on $\cal H$. Moreover, the vacuum is cyclic for the 
$\rho(w'[f'])$  In what follows we will drop the primes and the 
representation label $\rho$ in order to simplify the 
formulae.

\subsection{Constructing the OS measure from OS data}
\label{s2.1}

We consider the 
set of generating functionals of the {\it Wightman functions} (as $N$ varies)
\ba \label{2.2}
W((f_N,t_N),..,(f_1,t_1)) &:=&  
\langle\Omega,w(f_N,t_N)\;..\;w(f_1,t_1)\;\Omega\rangle_{{\cal H}} 
\nonumber\\
&=&
\langle\Omega,w(f_N)\;U(t_N-t_{N-1})^{-1}..U(t_2-t_1)^{-1}\;w(f_1)\;\Omega\rangle_{{\cal H}}  
\ea
If the labels $f$ correspond to semaring functions, then 
formally the actual Wightman functions are derived from (\ref{2.2}) 
by taking the functional 
derivatives with respect to the sharp time smearing functions 
$f_1,..,f_N$ and then setting them to zero. 

By definition, the {\it Schwinger functions} arise from the Wightman functions 
by analytic continuation in time
$t\mapsto i\beta,\; \beta\in \mathbb{R}$, provided, the continued times 
are ordered $\beta_N>\beta_{N-1}>..>\beta_1$ so that the unitary time evolution
operator becomes a self-adjoint contraction operator 
\be \label{2.3}
U(t_k-t_{k-1})^{-1}=\exp(i(t_k-t_{k-1})H)
\mapsto \exp(-(\beta_{k}-\beta_{k-1})H))
\ee
which is bounded in operator norm by unity thanks to the positivity of $H$.
Accordingly, we obtain the N-point Schwinger function generator without 
domain questions
\be \label{2.4}
S((f_N,\beta_N),..,(f_1,\beta_1)) := 
\langle\Omega,w(f_N)\; e^{-[\beta_N-\beta_{N-1}]H}..
e^{-[\beta_2-\beta_1]H}
w(f_1)\;\Omega\rangle_{{\cal H}}  
\ee
To define a path integral measure $\mu$ from (\ref{2.4}) we consider a 
spacetime field 
$\Phi$ taking values in some set $\Gamma$ to be specified (namely the 
support of $\mu$) and the cylindrical sharp time 
subsets of $\Gamma$ defined by
\be \label{2.5}
\Gamma^{\beta_1,..,\beta_N}_{B_1,..,B_N}=\{\Phi\in \Gamma;\;
\Phi(\beta_k)\in B_k;\; k=1,..,N\}
\ee
where $\beta_1<..<\beta_N$ are real numbers and $B_k$ are Borel sets 
with respect to the measure $\nu$ on $\Delta(\mathfrak{B})$ 
considered above. 
The set $\Gamma$ itself is also cylindrical and corresponds to the case 
$N=2,\; B_1=B_2=\Delta(\mathfrak{B})$ with any $\beta_1<\beta_2$ 
(no restriction). 
Then we assign 
to (\ref{2.5}) the ``heat kernel'' measure (heat kernel with respect to 
$H$)
\be \label{2.6}
\mu(\Gamma^{\beta_1,..,\beta_N}_{B_1,..,B_N}):=
\langle\Omega,\chi_{B_N} e^{-[\beta_N-\beta_{N-1}]H}..e^{-[\beta_2-\beta_1]H}
\chi_{B_1}\Omega\rangle
\ee
where $\chi_B$ is the operator that multiplies by $\chi_B(\phi),\;
\phi\in \Delta(\mathfrak{B})$ on the 
Hilbert space $L_2(\Delta(\mathfrak{B}),d\nu)$. Note that $\mu(\Gamma)=1$
due to $H\Omega=0$ and $\chi_{\Delta(\mathfrak{B})}=1_{{\cal H}}$.
The latter fact is responsible for the fact that our definition is 
cylindrically consistent (adding time slots with $1_{{\cal H}}$ insertions 
does not affect the answer). 
Apart from this, it is however unclear whether the numbers (\ref{2.6}),
although finite, are positive. If they are, they define the {\it Wiener
measure} with respect to $H$.

For a general theory it is a very difficult problem to decide whether (\ref{2.6}) is positive. 
In the literature, this property is known as Nelson-Symanzik positivity \cite{Sim74,FFS92}. 
While in general, (\ref{2.6}) need not be positive (cf. \cite{simon73} for a counter 
example), a more succinct criterion for its positivity is available.
Under the assumption of a vacuum $\Omega$ cyclic for $\mathfrak{B}$, the positivity can 
be reduced to the following positivity condition on $e^{-\beta H}$:
For any $F, G$ bounded, positive, measurable functions on 
the N-fold and M-fold copies of the ranges of the functions $w[f]$ (e.g.
$\mathbb C^N$ and $\mathbb C^M$ respectively for scalar fields)
and test functions $f_1,\ldots,f_N, g_1,\ldots,g_M$, it suffices to show that
\begin{equation} \label{NelsonSymanzik}
 \left<F(w(f_1),\ldots,w(f_N)), e^{-\beta H} G(w(g_1),\ldots,w(g_M))\right> \geq 0 \text{.}
\end{equation}
For completeness, we sketch the proof in appendix \ref{sc}.

In what follows we assume that the Wiener measure $\mu$ defined cylindrically
in (\ref{2.6}) exists. 
Then one can define the integral of simple functions 
(linear combinations of characteristic functions of cylindrical sets as in 
(\ref{2.5})) and approximate the integral of more complicated functions 
by simple functions, see e.g. \cite{Rud87} for the measure theoretic details
of this construction process. For (\ref{2.5}) 
we have 
\be \label{2.10}
\chi_{\Gamma^{\{\beta_k\}}_{\{B_k\}}}[\Phi]=
\prod_{k=1}^N \chi_{B_k}(\Phi(\beta_k))
\ee
and one sets 
\be \label{2.11}
\mu(\chi_{\Gamma^{\{\beta_k\}}_{\{B_k\}}}[\Phi]):=
\mu(\Gamma^{\{\beta_k\}}_{\{B_k\}})=
\mu(\prod_{k=1}^N \chi_{B_k}(\Phi(\beta_k))
\ee
In view of (\ref{2.4})
we are in particular interested in functions $G[\Phi]$
which depend on $\Phi$ only 
through $\Phi(\beta_k),\; k=1,..,N$ for a finite number of times. These can 
be approximated by a sequence $G_n$ of linear combinations of step functions 
as in (\ref{2.10}) for fixed $\beta_k$ but each $G_n$ involves many 
$B_k$. Concretely, one partitions $\Delta(\mathfrak{B})^N$ into mutually
disjoint cells of the 
form $B_1\times ..\times B_N$, picks a point in such a cell and 
approximates $G$ by the function that assumes the value of $G$ 
at this point throughout the whole cell. It is clear 
such an approximant is an (infinite) linear combination of functions 
(\ref{2.10}) where the sum is over all cells of the partition.
The integral $\mu(G)$, if it exists, is then defined by 
taking the limit of the integrals of the sequence\footnote{More in detail 
one does this first for positive functions, shows that there exists 
a sequence of simple functions smaller than $G$ pointwise a.e. wrt $\mu$
and then takes the supremum of the integral over all simple functions
that are smaller than $G$ a.e. wrt $\mu$. The integral for complex
valued functions is defined by decomposing real and imaginary part
in positive and negative contributions.}. 

Let now 
\be \label{2.12}
W[F]_{F=\delta_{\beta_k}\; f_k}:=w[f_k]_{\phi\to \Phi(\beta_k)} 
\ee
that is, $W[\delta_{\beta_k} f_k]$ is exactly built as the configuration field 
Weyl element $w[f_k]$, just that the time zero operator field 
$\phi(x),\;x\in \sigma$ 
is replaced by the history integration 
field $\Phi(\beta_k,x)$ at time $\beta_k$. Then 
due to the fundamental relation (\ref{2.6}) and what we have just said
about the approximation by simple functions we can define 
\be \label{2.13}
\mu(\prod_{k=1}^N\; W(\delta_{\beta_k} f_k))=
\langle\Omega,w(f_N)\;e^{-(\beta_N-\beta_{N-1})H}\; w(f_{N-1})\;...\;
e^{-(\beta_2-\beta_1)H}\; w(f_1)\Omega\rangle=S(\{f_k,\beta_k\})
\ee
Since the history field is just an integration variable, the Schwinger
function can be extended, using the measure $\mu$, 
to a function symmetric in all time variables 
in contrast to the Wightman function of course. That is why the sequence
of the product occurring on the left handside  of (\ref{2.13}) is irrelevant. 

Definition (\ref{2.13}) defines $\mu$ on the history smearing fields
of compact and sharp time support
\be \label{2.13a}
F:=(\delta_{\beta_k} \; f_k)_{k=1}^N,\;\;
W[F]:=\prod_k\; W[\delta_{\beta_k} f_k]
\ee
Note that we simply {\it define} $W[F]$ this way. For scalar fields, 
we may independently define $W[F]:=\exp(i\int\;d^{D+1}x F(x)\Phi(x))$ 
which is consistent with (\ref{2.13a}) but for more general theories
such as non-Abelian gauge theories for which $w[f]$ may for instance 
mean a Wilson loop variable $w[f]:={\rm Tr(Hol}_f(\phi))$ where $\phi$ is 
a connection, $f$ the distributional smearing function associated with a 
loop in $\sigma$ and Hol the holonomy map, the product of these Wilson 
loop variables cannot be expressed in terms of a time integral.  

Remark:\\
Equation (\ref{2.13}) holds for any $N$ and this is sufficient 
in order to reconstruct the Schwinger functions. One sometimes 
wishes to take the limit $N\to \infty$ 
in order to obtain a generating functional. 
To that end we consider a single function $F$ on 
$\mathbb{R}\times \sigma$ of compact time support in $[-\beta,\beta]$
and let $\beta_{k}=\beta\frac{k}{N},\; \beta>0,\; k=-N,-N+1,..,N$.
Then we set 
\be \label{2.13b}
f_k:=\int_{\beta_k-\beta/(2N)}^{\beta_k+\beta/(2N)}\; d\beta'\; F(\beta')
\ee
to be the average value of the single 
function $F$ over an interval of width $\beta/N$ around the 
sharp time $\beta_k$.
If we formally define 
\be \label{2.14}
W[F]:=\lim_{N\to \infty}\; \prod_{k=-N}^N\; W(\delta_{\beta_k} f_k)
\ee
one can try to give meaning to the limit
\be \label{2.15}
\mu(W[F])=\lim_{N\to \infty} 
\langle\Omega, w(f_N)\; e^{-\frac{\beta}{N} H}\; w(f_{N-1})\; .. \;
e^{-\frac{\beta}{N} H}\; w(f_{-N}) \Omega\rangle
\ee
For a scalar field (\ref{2.14}) converges to 
\be \label{2.16}
W[F,\Phi]=\exp(i \Phi[F]),\;\;\Phi[F]=\int_{\mathbb{R}\times \sigma}
d\beta \; d^Dx\; \Phi(\beta,x)\; F(\beta,x)
\ee
and $\mu(W[F])$, when it exists, is the generating functional of the Schwinger 
functions of a scalar field which one obtains by functional derivation 
with respect to $F$.
For a non-Abelian gauge theory such a simple formula is not available. That is why 
in this more general case we refer to (\ref{2.13}) as the Schwinger function.  
Notice that we recover 
(\ref{2.13}) from (\ref{2.15}) for the sharp time smearing functions of the 
form
\be \label{2.17}
F(\beta)=\sum_{k=1}^N\; f_k\; \delta(\beta,\beta_k)
\ee
so that (\ref{2.13}) can be considered as the more fundamental object while 
(\ref{2.15}) is a derived concept which is useful when it exists and 
then reproduces (\ref{2.13}). It transpires that
the measure is determined by $F$ with finite, discrete time support and 
we will deal only with that case in what follows as it applies also 
to quantum fields more general than scalar ones.\\
\\
We claim that $\mu$ so constructed is automatically time reflection 
invariant, time translation invariant and reflection positive.
To define these notions, let ${\cal H}':=
L_2(\Gamma,d\mu)$ be the history Hilbert space defined as the closed 
linear span of the functions $W[F]$ equipped with the scalar product
\be \label{2.18}
\langle W[F],W[F']\rangle_{{\cal H'}}:=\mu(\overline{W[F]} \;W[F'])
\ee
Notice that the Weyl relations (\ref{2.19}) are 
inherited by the functions $W[F]$ since the $W[F]$ are 
products of functions of the form $w[f]$ with $\phi$ replaced by 
$\Phi(\beta)$ for some $\beta$. That is, whenever $W[F],W[F']$ have 
coinciding points of time, the rules (\ref{2.19}) apply in 
$W[F]\; W[F']$ otherwise we just prolong the list of points of time.
In particular, if $F,F'$ have positive time support then we have 
$W[F] W[F']=W[\tilde{F}]$ for certain $\tilde{F}$ which still has positive 
time support. For a scalar field for instance $\overline{W[F]} W[F']=W[F'-F]$.

We define time reflection on the smearing functions and the history 
Hilbert space densely by 
\be \label{2.20}
(\theta F)(\beta):=F(-\beta),\;\; R\; W[F]:=W[\theta F]
\ee
We begin by showing that the measure $\mu$ is automatically reflection 
invariant. For $F(t)=(\delta_{\beta_k}(t) f_k)_{k=1}^N$ with 
$\beta_1<..<\beta_N$ we have 
$(\theta\cdot F)(t)=(\delta_{-\beta_k}(t) f_k)_{k=1}^N$ whence 
due to $-\beta_N<..<-\beta_1$
by definition of $\mu$
\ba \label{2.21}
\mu(R\cdot W[F]) 
&=& 
\langle\Omega, w[f_1]\; e^{-(-\beta_1-[-\beta_2])H}\; w[f_2]\;
e^{-(-\beta_2-[-\beta_3])H}\;..\; e^{-(\beta_{N-1}-[-\beta_N])H}\;
w[f_N]\;\Omega\rangle
\nonumber\\
&=& 
\langle\Omega, w[f_1]\; e^{-(\beta_2-\beta_1)H}\; w[f_2]\;
e^{-(\beta_3-\beta_2)H}\;..\; e^{-(\beta_N-\beta_{N-1})H}\;
w[f_N]\;\Omega\rangle
\nonumber\\
&=& 
\langle w[f_N]^\ast\; e^{-(\beta_N-\beta_{N-1})H}\;..\; e^{-(\beta_2-\beta_1)H}\;
w[f_1]^\ast\;\Omega,\Omega\rangle
\nonumber\\
&=&
\overline{\langle \Omega, 
w[f_N^\ast]\; e^{-(\beta_N-\beta_{N-1})H}\;..\; e^{-(\beta_2-\beta_1)H}\;
w[f_1^\ast]\Omega\rangle }
\nonumber\\
&=&
\overline{\mu(W[F^\ast])}
=\int_\Gamma\; d\mu(\Phi)\; \overline{W[F^\ast]} 
\nonumber\\
&=& \mu(W[F^\ast]^\ast)
=\mu(W[F])
\ea
where positivity of the measure was used as well as involutivity of $^\ast$.
It follows that $R$ is a unitary
operator on ${\cal H}'$  due to $R^2=1_{{\cal H}'}$.

We consider the subspace $V$ of ${\cal H}'$ consisting of the finite 
linear span of the $W[F]$ where $F$ has positive finite and discrete 
time support in 
$(0,\infty)$. Then $\mu$ is said to be reflection positive iff
\be \label{2.21a}
\langle \Psi,R\Psi\rangle_{{\cal H}'}\ge 0\;\;\forall\;\; \Psi\in V
\ee
We consider an arbitrary element of $V$ which can be written as
\be \label{2.22}
\Psi=\sum_{I=1}^M\; z_I\; W[F^I],\;\;{\rm T-supp}(F^I)\subset (0,\infty)
\ee
and work out condition (\ref{2.21a}). The crucial observation is that 
if T-supp$(F)\subset (0,\infty)$ then 
T-supp$(\theta F)\subset (-\infty,0)$. Suppose that 
$F^I=\sum_{k=1}^{N^I} \delta_{\beta^I_k} f^I_k,\; 
0<\beta_1^I<..<\beta_{N^I}^I$ then similarly as above 
\ba \label{2.23}
 \langle\Psi,R\Psi\rangle
&=& \sum_{I,J}\; \bar{z}_I\; z_J\;\mu(\overline{W[F]}\; W[\theta\cdot F])
\nonumber\\
&=& \sum_{I,J}\; \bar{z}_I\; z_J\;\mu(W[F^\ast]\; W[\theta\cdot F])
\nonumber\\
&=& \sum_{I,J}\; \bar{z}_I\; z_J\;
\langle \Omega,w[(f^I_{N^I})^\ast]\; e^{-(\beta^I_{N^I}-\beta^I_{N^I-1})H}\;..\;
e^{-(\beta^I_2-\beta^I_1)H} w[(f^I_1)^\ast]\times
\nonumber\\
&& \times
e^{-(\beta^I_1-[-\beta^J_1])H}\; w[f^J_1]\; e^{-(\beta^J_2-\beta^J_1)H}\;
..\; e^{-(\beta^J_{N^J}-\beta^J_{N^J-1})H}\; w[f^J_{N^J}]\; \Omega\rangle
\nonumber\\
&=& \sum_{I,J}\; \bar{z}_I\; z_J\;
\langle e^{-\beta^I_1 H}\; w[f^I_1]\; e^{-(\beta^I_2-\beta^I_1)H}\;
..\; e^{-(\beta^I_{N^I}-\beta^I_{N^I-1})H}\; w[f^I_{N^I}]\; \Omega,\;
\nonumber\\
&&\;\;\;\;
e^{-\beta^J_1 H}\; w[f^J_1]\; e^{-(\beta^J_2-\beta^J_1)H}\;
..\; e^{-(\beta^J_{N^J}-\beta^J_{N^J-1})H}\; w[f^J_{N^J}]\; \Omega\rangle 
\nonumber\\
&=& 
|| \sum_I\; z_I\;e^{-\beta^I_1 H}\; w[f^I_1]\; e^{-(\beta^I_2-\beta^I_1)H}\;
..\; e^{-(\beta^I_{N^I}-\beta^I_{N^I-1})H}\; w[f^I_{N^I}]\; \Omega||^2
\ea
which is manifestly non-negative. 

Finally, we show that $\mu$ is trivially time translation invariant. We 
define densely on ${\cal H}'$ 
\be \label{2.23a}
(T_s F)(\beta'):=F(\beta'-s),\;\; {\cal U}(s)W[F]:=W[T_s F]
\ee
Thus for $F=(\delta_{\beta_k}\; f_k)_{k=1}^N,\;\beta_1<..<\beta_k$ we have 
$T_s\cdot F=(\delta_{\beta_k+s}\;f_k)_{k=1}^N$ hence with $\beta'_k=\beta_k+s$
we still have $\beta'_1<..<\beta'_N$ so that
\be \label{2.23b}
\mu({\cal U}(s)\ W[F])
=\langle \Omega,w[f_N] e^{-(\beta'_N-\beta'_{N-1})H}\;..\; 
e^{-(\beta'_2-\beta'_1)H}\;w[f_1]\;\Omega \rangle
=\mu(W[F])
\ee
since $\beta'_k-\beta'_l=\beta_k-\beta_l$.
Note that $s\mapsto 
{\cal U}(s)$ is in particular a one-parameter unitary group on ${\cal H}'$.
Now the strong continuity
of the one parameter unitary group $t\mapsto U(t)=e^{-it H}$ on $\cal H$, 
equivalent
to the existence of the self-adjoint generator $H$ by Stone's theorem, implies
strong continuity of the one-parameter semi-group $\beta\mapsto e^{-\beta H}$
with $\beta>0$ on $\cal H$. This in turn implies the strong continuity of 
the unitary group ${\cal U}(s)$. As we will see in the next section, 
the bi-linear form (\ref{2.21a}) is positive but not definite and thus 
has a null space which leads to corresponding equivalence classes $[\Psi],\;
\Psi\in V$. The space $V$ is only invariant under ${\cal U}(s)$ for $s\ge 0$
and thus gives rise to a semigroup $K'(s)[\Psi]:=[{\cal U}(s)\Psi]$ (see the next section 
for proving independence of the representative). It follows 
that the semigroup $s\mapsto K'(s)$ is also strongly continuous, a property
that maybe called ``reflection continuity'' and which implies the existence 
of a generator of the semigroup which gives rise to a Hamiltonian defined 
within the measure theoretic framework, see the next subsection.\\
\\
In conclusion, we have shown that the Hamiltonian framework, under the 
assumption of positivity of (\ref{2.6}) and under the assumptions made in the 
preliminaries, always gives rise to a measure on an appropriate 
history field space $\Gamma$ which is time translation and time reflection 
invariant as well as reflection positive and continuous. 

We will not require the remaining Osterwalder-Schrader axioms, that is,
Euclidian invariance of the measure, analyticity of the generating functional 
in the smearing fields $F$, regularity as an estimate of the 
generating functional in terms of $L_p$ norms of $F$ and 
clustering\footnote{Alternatively: ergodicity of the 
measure with respect to time translations, that is, time average 
of $\Psi\in L_1(\mu)$ and ensemble average $\mu(\psi)$ coincide $\mu$ a.e.}
of the measure with respect to time translations.
Euclidian invariance maybe too strong to ask for our theory under 
investigation (e.g. General Relativity), analyticity maybe too strong to 
ask if the state $\omega$ is not regular for $\mathfrak{B}$
(i.e. not 
strongly continuous with respect to $F$ in $W[F]$) in which case also the 
regularity OS axiom will fail to hold and finally ergodicity, which is 
equivalent to the uniqueness of the vacuum of $H$ cannot be verified here 
unless we put this in from the outset into the canonical theory, which we do
not want to do because in more general theories, the vacuum sector maybe
degenerate (phase transitions).

\subsection{Constructing the OS data form the OS measure}
\label{s2.2}

The converse statement is standard, we include it here only for completeness,
to have a unified notation 
and in order to draw attention to the fact that only reflection invariance 
and reflection positivity as well as time translation invariance and 
continuity are needed in order to construct the OS data.
Otherwise we will be brief, see \cite{GJ87} for all the details in a Minkowski
space context and \cite{AMMT99} for the context in a generally covariant theory.\\
\\
We assume to be given a time reflection and time translation invariant as well
as reflection positive and continuous measure on the history field space 
$\Gamma$. 
By definition of reflection positivity, the bi-linear form on $V$, which
is the space of
the finite linear combinations of the $W[F]$ with $F$ of strictly 
positive time support, given by
\be \label{2.24}
(\Psi,\Psi')_V:=\langle \Psi,R\;\Psi'\rangle_{{\cal H}'}
\ee
is positive but not necessarily positive definite on $V$. However, due to 
reflection invariance, in particular unitarity of the 
time reflection operator $R$, it is actually sesqui-linear
\be \label{2.24a}
\overline{(\Psi,\Psi')_V}=
\langle R\;\Psi',\Psi\rangle_{{\cal H}'}
=\langle R^2\;\Psi',R\; \Psi\rangle_{{\cal H}'}=(\Psi',\Psi)_V
\ee
 
We consider the null space 
\be \label{2.25}
{\cal N}:=\{\Psi\in V;\;(\Psi,\Psi)=0\}
\ee
and define the canonical Hilbert space ${\cal H}$ as the completion
of the equivalence classes $[\Psi]:=\{\Psi+\Psi_0,\; \Psi_0\in {\cal N}\}$
with $\Psi\in V$
with respect to the scalar product
\be \label{2.25a}
\langle [\Psi],[\Psi']\rangle_{{\cal H}}:=(\Psi,\Psi')_V
\ee
This definition is independent of the representative by virtue of the 
Cauchy-Schwarz inequality which applies to positive semi-definite 
sesquilinear forms. Note that T-supp$(T_s F)$=T-supp$(F)+s$, thus the 
unitary operator densely defined on ${\cal H}'$ by ${\cal U}(s) W[F]
=W[T_s F]$ does not map $V$ onto $V$ unless $s>0$. Thus, on $V$ we may still define 
a one parameter family $s\mapsto {\cal U}(s)$, but it is a semi-group
rather than a group
because the inverses are not defined. We want to define for  
$\Psi\in V$
\be \label{2.26}
K(s) [\Psi]:=[{\cal U}(s)\Psi]  
\ee
but must show that the definition is independent of the representative, i.e.
that ${\cal U}(s) {\cal N}\subset {\cal N}$. We have 
\be \label{2.27}
\theta\circ T_s \cdot F(\beta)=(T_s\cdot F)(-\beta)=F(-\beta-s)=
F(-(\beta+s))=(\theta \cdot F)(\beta+s)=T_{-s}\circ \theta \cdot F(\beta)
\ee
whence
\be \label{2.28}
R\; {\cal U}(s)\; W[F]=W[\theta \circ T_s F]={\cal U}(-s)\; R\; W[F]
\ee
We therefore have, using unitarity
of ${\cal U}(s)$ on ${\cal H}'$, for $\Psi\in V,\; \Psi_0\in {\cal N}$
\ba \label{2.29}
|(\Psi,{\cal U}(s)\; \Psi_0)| &=&
|\langle\Psi,R {\cal U}(s) \Psi_0\rangle_{{\cal H}'}|=
|\langle\Psi,{\cal U}(-s) R\;\Psi_0)\rangle_{{\cal H}'}|=
\\
&=&
|\langle {\cal U}(s) \Psi,R\;\Psi_0)\rangle_{{\cal H}'}|=
|({\cal U}(s) \Psi,\Psi_0))_V|\le 
||{\cal U}(s) \Psi||_V\;\; ||\Psi_0||_V=0
\nonumber
\ea
The same calculation as in (\ref{2.29}) demonstrates that $K(s)$ is symmetric
on ${\cal H}$ and positive since 
\be \label{2.30}
\langle [\Psi], K(s) [\Psi]\rangle_{{\cal H}}= 
\langle [\Psi], [{\cal U}(s/2)^2 \Psi]\rangle_{{\cal H}} =
\langle [\Psi], K(s/2)^2 [\Psi]\rangle_{{\cal H}} =
||K(s/2)\; [\Psi]||_{{\cal H}}^2
\ee
We show that $K(s)$ is a contraction. Consider any 
$0\not=[\Psi]\in {\cal H}$ (in particular $0\not=\Psi\in {\cal H}'$),
then using the Cauchy-Schwarz inequality for $\cal H$ repeatedly
\ba \label{2.31}
||K(s) [\Psi]||_{{\cal H}}
&=& \sqrt{\langle [\Psi],K(s)^2 [\Psi]\rangle_{{\cal H}} }
\nonumber\\
&\le & ||[\Psi]||^{1/2}\;\; [||K(2s) [\Psi] ||_{{\cal H}}]^{1/2}  
\nonumber\\
&\le & ||[\Psi]||^{\sum_{k=1}^n 2^{-k}}\;\;
[||K(2^n s) [\Psi] ||_{{\cal H}}]^{2^{-n}}  
\ea
Now for any $r>0$, using the unitarity of $R, {\cal U}(r)$ for ${\cal H}'$
and its Cauchy-Schwarz inequality 
\be \label{2.32}
||K(r)[\Psi]||_{{\cal H}}^2
=||[{\cal U}(r)\Psi]||_{{\cal H}}^2
=\langle {\cal U}(r)\Psi, R {\cal U}(r) \Psi\rangle_{{\cal H}'}
\le ||{\cal U}(r)\Psi||_{{\cal H}'} \;
||R {\cal U}(r)\Psi||_{{\cal H}'} \le 
||\Psi||^2_{{\cal H}'}
\ee
Thus, for any $n$ 
\be \label{2.33}
||K(s) [\Psi]||_{{\cal H}} \le  ||[\Psi]||_{{\cal H}}^{1-2^{-n}} \;\;
||\Psi||_{{\cal H}'}^{2^{-n}}
\ee
and taking the limit $n\to \infty$ 
\be \label{2.34}
\frac{||K(s) [\Psi]||_{{\cal H}}}{||[\Psi]||_{{\cal H}}}
\le 1
\ee
Taking the supremum over $[\Psi]\not=0$ shows that $K(s)$ is bounded in norm
by one, i.e. $s\to K(s)$ is a contraction semigroup on ${\cal H}$. 
Note that there are two Hilbert space norms involved in (\ref{2.33}) 
so that unitarity of ${\cal U}(r)$ alone would not suffice to establish
(\ref{2.34}).

Next,
by assumption ${\cal U}(s)$ is reflection continuous whence for $\Psi\in 
{\cal H}$
\ba \label{2.33a}
||[K(s)-1_{{\cal H}}][\Psi]||^2
&=& \langle \Psi,[{\cal U}(2s)-2{\cal U}(s)+1_{{\cal H}'}]R\Psi\rangle_{\mathcal{H}'}
\\
&\le& |\langle\Psi,[{\cal U}(2s)-1_{{\cal H}'}] R\Psi\rangle_{{\cal H}'}|
+2 |\langle\Psi, [{\cal U}(s)-1_{{\cal H}'}]R\Psi\rangle_{\mathcal{H}'}|\nonumber
\ea
tends to zero as $s\to 0$. Hence the contraction semi-group is 
continuous and therefore has a self-adjoint generator $H$ by the Hille-Yosida 
theorem \cite{RS80}. 

Finally, we define $\Omega:=[1]$ where $1\in {\cal H}'$ is the constant
function equal to unity which we may think of as $W[F]_{F=0}$. Then 
evidently
\be \label{2.35}
K(s) \Omega=[{\cal U}(s) W[F]]_{F=0}]=[W[T_s F]_{F=0}]=[1]=\Omega
\ee
so that $\Omega$ is a vacuum for $H$.

For the sake of completeness, we show that the vacuum is unique if 
if $\mu$ is time-clustering, i.e.
\be \label{2.36}
\lim_{s\to \infty} \; \mu(W[F]^\ast \;W[T_s F'])=
\mu(W[F]^\ast) \; \mu(W[F'])=\langle W[F],1\rangle_{{\cal H}'} \; \langle 1,W[F']\rangle_{{\cal H}'}
\ee
Using that the span of the $W[F]$ is dense, 
property (\ref{2.36}) says that, as $s\to \infty$, 
the operator ${\cal U}(s)$ becomes the projection operator 
$P=1\; \langle 1,.\rangle_{{\cal H}'}$ on ${\cal H}'$ in the weak 
operator topology. Now suppose that 
$K(s)\Omega=\Omega$. We find $\Omega'\in {\cal H}'$ such that 
$[\Omega']=\Omega$. It follows 
\be \label{2.37}
[K(s)-1_{{\cal H}}]\Omega=[[{\cal U}(s)-1_{{\cal H}'}]\Omega']=0
\ee
i.e. $[{\cal U}(s)-1_{{\cal H}'}]\Omega' \in {\cal N}$ for any $s$. Thus,
if $\mu$ clusters, then by taking the limit in (\ref{2.37}) we find 
that $\langle 1,\Omega'\rangle 1-\Omega'\in {\cal N}$, i.e. $\Omega:=[\Omega']\propto
[1]$ and the vacuum is unique (up to a phase when normalised).\\
\\
Remark:\\
The Osterwalder-Schrader reconstruction of the data $({\cal H},\Omega,H)$ just relies
on the time reflection and translation invariance, time continuity and time reflection
positivity. Neither is anything being said about the existence of a ``mass gap'' nor 
does OS reconstruction rely on it, not even when restricting to QFT 
in Minkowski space. For QFT in Minkowski space, by a mass gap is meant a 
strictly positive infimum in the support, without zero, of the 
K\"allen-Lehmann spectral measure for the 2-point function and as such is a very 
``Minkowskian'' concept. Such a mass gap is 
important in order to develop Haag-Ruelle scattering theory (rigorous collision theory)
and in particular for constructive Yang-Mills theory it would be an indirect 
proof of confinement. In our general theory setting, which in principle 
encompasses quantum gravity in suitable gauges, it is not even expected that
the 2-point function is related in such a simple way to a suitable notion 
of mass since the quantum gravity vacuum would presumably describe a spacetime with 
a degenerate geometry 
expectation value. Hence, such and related questions can meaningfully be asked only 
when adding further structural elements (e.g. correlation functions in states 
describing a semiclassical geometry).

\subsection{OS data reconstruction and OS measure construction are inverses}
\label{s2.3}

We wish to demonstrate that OS construction (i.e. computing the automatically
reflection positive Wiener 
measure from OS data) and OS reconstruction (i.e. computing OS data 
from a reflection positive measure) are inverse processes. Recall from our 
preliminary remarks that  
is implicit in our assumptions that, if necessary, we have recast our 
formulation into a form 
such that when starting with the OS data,
the sharp time zero Weyl elements $w[f]$ are well defined and 
the $w[f]\Omega$ lie dense in the 
canonical Hilbert space ${\cal H}$. Likewise, starting with the OS measure, 
if the sharp time $W[\delta_0 f]$ are not well defined as $\mu$ measurable 
and integrable functions and/or 
the $[W[\delta_0 f]]$ do not lie dense in the OS Hilbert space, we make 
use of our assumption spelled out in the preliminaries that the OS Hilbert 
space $\cal H$ admits an Abelian C$^\ast-$subalgebra $\mathfrak{B}$ of 
${\cal B}({\cal H})$ such that $\mathfrak{B}\Omega$ is dense. We then 
reformulate the OS measure $\mu$ in terms of the Weyl elements $w[f]$ 
generating $\mathfrak{B}$. Then the corresponding functions of the 
history field $\Phi$ given by    
$W[\delta_s f]_{\Phi}=w[f]_{\phi\to \Phi(s)}$ exist as $\mu-$integrable 
functions and allow to construct the generating functional of 
Schwinger functions. It is also being understood that we have reformulated 
$\mu$ in  this way, if necessary.

\subsubsection{Reproduction of OS data}
\label{s2.3.1}

We first start with OS data $({\cal H},\Omega, H)$ and construct the 
Wiener measure 
$\mu$ out of them according to section \ref{s2.1}. We ask, whether
the reconstruction algorithm of section \ref{2.2} recovers these data.

Concerning the null space, using the same notation as in (\ref{2.22}), 
(\ref{2.23}), recall 
\be \label{2.42}
||[\Psi]||_{\overline{V/{\cal N}}}^2
=
|| \sum_I\; z_I\;e^{-\beta^I_1 H}\; w[f^I_1]\; e^{-(\beta^I_2-\beta^I_1)H}\;
..\; e^{-(\beta^I_{N^I}-\beta^I_{N^I-1})H}\; w[f^I_{N^I}]\; 
\Omega||_{{\cal H}}^2
\ee
Consider the vector in $\cal H$
\be \label{2.43}
\psi:=\sum_I\; z_I\;e^{-\beta^I_1 H}\; w[f^I_1]\; e^{-(\beta^I_2-\beta^I_1)H}\;
..\; e^{-(\beta^I_{N^I}-\beta^I_{N^I-1})H}\; w[f^I_{N^I}]\; 
\Omega
\ee
that appears in (\ref{2.42}). Since by assumption 1. above 
the finite linear span of 
the $w[f]\Omega$ lies dense in 
${\cal H}$, for any $\epsilon>0$ we find $\psi^\epsilon=\sum_J\; c_J\;
w[g^J]\Omega$ such that $||\psi-\psi^\epsilon||_{{\cal H}}<\epsilon$. 
Consider the corresponding $\Psi^\epsilon=\sum_J c_J W[G^J],\;
G^J=\delta_0 g^J$. Then by the same calculation as in (\ref{2.23})
\be \label{2.43a}
||[\Psi]-[\Psi^\epsilon]||_{\overline{V/{\cal N}}}^2=
||\psi-\psi^\epsilon||_{{\cal H}}^2<\epsilon^2
\ee
Since the scalar product on $\cal H$ is non-degenerate,
this demonstrates that the equivalence class of $[\Psi]$ can be 
labelled by representatives which lie in the closure of the span of 
the $W[F]$ with $F=\delta_0 f$. This suggests to define the densely 
defined embedding
\be \label{2.43b}
E:\; {\cal H}\to V,\;\psi=\sum_I\; z_I\; w[f^I]\;\Omega \mapsto
\Psi=\sum_I \; z_I\; W[\delta_0 f^I]
\ee
It follows that 
$E(\Omega)=1$ is the constant function equal to unity and the scalar 
products are isometric $||[E(\psi)]||_{\overline{V/{\cal N}}}=
||\psi||_{\cal H}$, i.e. $[E(\psi)]$ can be identified with 
$\psi$. This shows that the 
equivalence classes of vectors with respect to the reflection positive 
inner product are precisely labelled by vectors in $\cal H$ and $E$ defines a 
section in the Hilbert bundle $\pi: V\to {\cal H}$ with projection 
$\pi(.)=[.]$, that is, $\pi\circ E={\rm id}_{{\cal H}}$ and $V,{\cal H}$ 
are respectively total space and base of the bundle. We conclude 
$\overline{V/{\cal N}}\equiv {\cal H}$.         

It follows then from (\ref{2.42}) and (\ref{2.43})
that for $F=(\delta_{\beta_k} f_k)_{k=1}^N,\; 0<\beta_1<..<\beta_N$ 
\be \label{2.44}
[W[F]]\equiv e^{-\beta_1 H}\; w[f_1]\; e^{-(\beta_2-\beta_1)H}\; w[f_2]\;..\;
e^{-(\beta_N-\beta_{N-1})H}\; w[f_N]\;\Omega
\ee
We compute the contraction semi-group for 
$\psi=\sum_I\; z_I\; w[f^I]\;\Omega$ from (\ref{2.44}),
\begin{align} \label{2.45}
K(s)\psi &\equiv & K(s)\; [E(\psi)]=[{\cal U}(s) E(\psi)]=
[\sum_I\; z_I\; W[\delta_s f^I]]  
\equiv \sum_I\; z_I\; e^{-s H}\; w[f^I]\; \Omega  
=e^{-s H}\psi
\end{align}
whence $K(s)=e^{-sH}$ indeed.\\
\\
In conclusion, we have shown that the Wiener measure constructed from the 
data $({\cal H},\Omega,H)$ has precisely those data as OS data. 

\subsubsection{Reproduction of the OS measure}
\label{s2.3.2}

Conversely, suppose that we are given an OS measure $\mu$. Does the 
Wiener measure $\mu'$ constructed from its OS data 
$({\cal H}=\overline{V/{\cal N}},\Omega=[1],H=-[d/ds]_{s=0} K(s))$ 
coincide with the measure $\mu$ that we started from? To define $\mu'$
we must actually first {\it define} the Weyl operators $w[f]$ on the 
Hilbert space $\cal H$. According to introduction to this subsection, it 
is meaningful to define them densely  
($G=(\delta_{\beta_k} g_k)_{k=1}^N,\;0<\beta_1,..,\beta_N$ is of discrete 
and positive time support) by 
\be \label{2.46a}
w[f] [W[G]]:= [W[F^0_f] W[G]],\;\;F^0_f:=\delta_0 f
\ee
In general, we set $F^s_f:=\delta_s f$. Notice the 
identities $T_r F^s_f=F^{s+r}_f, \; \theta F^s_f=F^{-s}_f$. 
Recall also $K(s)[\Psi]=[{\cal U}(s)\Psi]$.
Definition (\ref{2.46a}) is independent of the representative because 
suppose that 
$[\Psi]=0$ with $\Psi$ in the finite linear span of the $W[F]$ with 
$F$ of positive time support then 
\ba \label{2.46b}
|| w[f]\;[\Psi]||_{{\cal H}}^2 
&=&\langle W[F^0_f]\; \Psi,\; R\cdot (W[F^0_f] \Psi)\rangle_{{\cal H}'}
=\langle W[F^0_f]\; \Psi,\; W[F^0_f]\; (R\cdot \Psi)\rangle_{{\cal H}'}
\nonumber\\
&=& \langle w[f]^\ast\; w[f] \;[\Psi],\; [\Psi]\rangle_{{\cal H}}
\le ||w[f]^\ast\; w[f] \;[\Psi]||_{{\cal H}}\;\;||[\Psi]||_{{\cal H}} 
=0
\ea
by the Cauchy-Schwarz inequality. 
We used that $\theta \cdot \delta_0=\delta_0$. Indeed this was very crucial 
because if representing $w[f]$ in terms of some $W[F]$ had also involved strictly 
positive time support in $F$ then 
we could not have made this conclusion because $R W[F]=W[\theta F] R$ and 
$W[\theta F]^\ast$ does not even preserve $V$.

We compute for $F$ of discrete time support at $\beta_1<..<\beta_N$
and with $F^s_k:=F^s_{f_k}$ 
\ba \label{2.46}
\mu'(W[F]) &=&
\langle \Omega,w[f_N]\; K(\beta_N-\beta_{N-1})..K(\beta_2-\beta_1) w[f_1] 
\Omega\rangle_{{\cal H}}
\nonumber\\
&=&    
\langle \Omega,w[f_N]\; K(\beta_N-\beta_{N-1})..K(\beta_2-\beta_1) 
[W[F^0_1]\cdot 1]\rangle_{{\cal H}}
\nonumber\\
&=&    
\langle\Omega,w[f_N]\; K(\beta_N-\beta_{N-1})..K(\beta_3-\beta_2)
w[f_2]\; 
[{\cal U}(\beta_2-\beta_1)\; W[F^0_1]\cdot 1]\rangle_{{\cal H}}
\nonumber\\
&=&    
\langle \Omega,w[f_N]\; K(\beta_N-\beta_{N-1})..K(\beta_3-\beta_2)
w[f_2]\; 
[W[F^{\beta_2-\beta_1}_1]]\rangle_{{\cal H}}
\nonumber\\
&=&    
\langle\Omega,w[f_N]\; K(\beta_N-\beta_{N-1})..K(\beta_3-\beta_2)\; 
[W[F^0_2]\; W[F^{\beta_2-\beta_1}_1]]\rangle_{{\cal H}}
\nonumber\\
&=&    
\langle \Omega,w[f_N]\; K(\beta_N-\beta_{N-1})..K(\beta_4-\beta_3)\; 
w[f_3]\;
[W[F^{\beta_3-\beta_2}_2]\; W[F^{\beta_3-\beta_1}_1]]\rangle_{{\cal H}}
\nonumber\\
&=&    
\langle [1],[W[F^{\beta_N-\beta_N}_N]\;...\;W[F^{\beta_N-\beta_1}_1]]\rangle_{{\cal H}}
\nonumber\\
&=&    
\langle 1,R W[F^{\beta_N-\beta_N}_N]\;...\;W[F^{\beta_N-\beta_1}_1]\rangle_{{\cal H}'}
\nonumber\\
&=&    
\langle 1,W[F^{\beta_N-\beta_N}_N]\;...\;W[F^{\beta_1-\beta_N}_1]\rangle_{{\cal H}'}
\nonumber\\
&=&    
\langle 1,{\cal U}(-\beta_N)\;W[F^{\beta_N}_N]\;...\;W[F^{\beta_1}_1]\rangle_{{\cal H}'}
\nonumber\\
&=&    
\langle {\cal U}(\beta_N)\cdot 1,\;
W[F^{\beta_N}_N]\;...\;W[F^{\beta_1}_1]\rangle_{{\cal H}'}
\nonumber\\
&=&    
\langle 1,\;W[F]\rangle_{{\cal H}'}=\mu(W[F])
\ea
thus indeed the Wiener measure coincides with $\mu$.

\section{Brief review of path integral renormalisation}
\label{s3}

In constructive QFT one works in the Euclidian framework and seeks 
to construct a measure $\mu$ on the history space of 
fields $\Phi\in \Gamma$ satisfying the Osterwalder-Schrader axioms.
The construction uses both an infrared (IR) cut-off $R$ and an 
ultraviolet (UV) cut-off $\epsilon$. In principle, one can distinguish between
spatial cut-offs $R,\epsilon$ and temporal cut-offs $T,\delta$ although 
one often identifies them $\epsilon=\delta$ and $T=R$. We will keep
the distinction, because in the Hamiltonian framework, time is always 
a continuous parameter and therefore is treated differently. Making 
the distinction between spatial and temporal will enable us to simplify
the match between
the path integral and Hamiltonian renormalisation later on.\\
\\ 
In order to understand the role of $T,\delta,R,\epsilon$ and how 
they find their way into the Euclidian formulation, we start from the 
observation, that the measure $\mu$ is completely specified by its 
generating functional $\mu(W[F])$ where $W[F,\Phi]$ involves the 
smearing field $F$ of the history quantum field $\Phi$. The IR cut-offs
are introduced simply by restricting the support of $F$ to $[0,T]\times
\sigma_R$ where $\sigma_R$ is a compact submanifold of $\sigma$, for
instance $[0,R]^D$ if $\sigma=\mathbb{R}^D$. Thus, such 
smearing functions probe $\Phi$ only on a compact submanifold of
$M=\mathbb{R}\times \sigma$. The UV cut-offs are introduced by restricting 
to smearing functions that in addition  
probe $\Phi$ only with finite temporal and spatial resolution $\delta$ and 
$\epsilon$ respectively. For instance, such a
function could be constant on each cell of a lattice in $[-T,T]\times
\sigma_R$ with cell size of the order of $\delta \epsilon^D$ and thus 
is completely specified by its value on the lattice vertices. 
In what follows, we will trade the numbers $\delta,\epsilon$ 
for the numbers of $N,M$ which indicate the number of vertices of the 
lattice in the temporal and each spatial direction respectively. 
This means $\delta=\delta_{T,N}=\frac{T}{N}, \; \epsilon=\epsilon_{R,M}
=\frac{R}{M}$ 
if we consider a regular hypercubic lattice in $\mathbb{R}^{D+1}$.
In general, the relation among the $T,N,\delta$ and the $R,M,\epsilon$
maybe more involved. 

To have 
something concrete in mind, for a scalar field on $\mathbb{R}^{D+1}$
such a compactly supported function of the indicated resolution
could be given by
\ba \label{3.1}
F_{T,N,R,M}(t,x) &=& \sum_{n\in \mathbb{Z}_N,\;m\in \mathbb{Z}_M^D}
\hat{F}_{T,N,R,M}(n,m)\; 
\chi_{n\delta_{T,N},m\epsilon_{R,M}}(t,x)
\nonumber\\
\chi_{n\delta_{T,N},m\epsilon_{R,M}}(t,x)
&=&\chi_{[n\delta_{T,N},(n+1)\delta_{T,N})}(t)
\prod_{a=1}^D
\;\chi_{[m^a\epsilon_{R,M},(m^a+1)\epsilon_{R,M})}(x^a)
\ea
where $\mathbb{Z}_N:=
\{0,1,..,N-1\}
,\;\mathbb{Z}_M:=
\{0,1,..,M-1\}$.
Note that indeed the coefficient 
$\hat{F}_{T,N,R,M}(n,m)$ in (\ref{3.1}) {\it is} the 
value of $F_{T,N,R,M}(t,x)$ at $(t,x)=(n\delta_{T,N},m \epsilon_{R,M})$.

We want to distinguish between the continuum function $F_{T,N,R,M}$ 
and its restriction to the lattice parametrised by $T,N,R,M$. 
Let $L_{T,R}$ be the space of compactly supported 
continuum smearing functions and 
$L_{T,N,R,M}$ the space of lattice smearing functions. In the 
example above, $L_{T,N,R,M}$ is simply the set of finite sequences
$\hat{F}_{T,N,R,M}(n,m),\; n\in \mathbb{Z}_N,\; m\in \mathbb{Z}_M^D$. 
Then we obtain maps
\begin{align} \label{3.2}
E_{T,N,R,M}:\; &L_{T,R}\to L_{T,N,R,M}\nonumber\\
 &(E_{T,N,R,M}\cdot 
F_{T,R})(n,m):=
F_{T,R}(n\delta_{T,N},m\epsilon_{R,M})
\nonumber\\
I_{T,N,R,M}:\; &L_{T,N,R,M}\to L_{T,R}\nonumber\\ 
&(I_{T,N,R,M} \hat{F}_{T,N,R,M})(t,x):=
\sum_{n\in \mathbb{Z}_N,\;m\in \mathbb{Z}_M^D}
\hat{F}_{T,N,R,M}(n,m)\; 
\chi_{n\delta_{T,N},m\epsilon_{R,M}}(t,x)
\end{align}
In fact, in this concrete situation the map $E_{T,N,R,M}$ is redundant
because the spaces $L_{T,N,R,M},\; L_{T,R}$ here carry the additional 
structure of an inner product so that instead of $E_{T,N,R,M}$ we can 
consider $I_{T,N,R,M}^\dagger$. This has the advantage 
of introducing less structure at the price of keeping the inner products.
We keep $E_{T,N,R,M}$ as a separate entity for more general situations in 
which natural inner products are not available. 

Note that while the {\it evaluation map} $E_{T,N,R,M}$ is canonical and 
theory independent (it just requires the knowledge of the relation
between the integers $(n,m)$ and the points of the lattice), 
the {\it injection map} $I_{T,N,R,M}$ 
is not canonical and does depend on the type of the theory under 
consideration and is of the above form only for a scalar field. For 
other types of fields it will assume a different form, but still such 
a map can always be invented. We will restrict the choice of 
$I_{T,N,R,M}$ by insisting that 
\be \label{3.2a}
E_{T,N,R,M} \circ I_{T,N,R,M}={\rm id}_{L_{T,N,R,M}}
\ee
which is satisfied for the choice (\ref{3.2}). It means that the
discrete information contained in a lattice function can be fully
recovered by evaluation after injection.
It will be the choice of the injection map that will determine the coarse 
graining map and therefore the whole renormalisation procedure. 
As it happens, for the concrete choice (\ref{3.2}) condition (\ref{3.2a})
is also satisfied if we replace $I_{T,N,R,M}^\dagger$ for $E_{T,N,R,M}$.
This means that $I_{T,N,R,M}$ is there an {\it isometric} embedding. 

The formula that defines $W[F_{T,R},\Phi_{T,R}]$, that is, the pairing between 
$F_{T,R}\in L_{T,R}$ and $\Phi_{T,R}\in \Gamma_{T,R}$  
in the continuum induces a discretised
lattice field $\hat{\Phi}_{T,N,R,M}$ and a corresponding pairing via
\be \label{3.4}
W_{T,R}[I_{T,N,R,M} \hat{F}_{T,N,R,M},\Phi_{T,R}]=:
W_{T,N,R,M}[\hat{F}_{T,N,R,M},\hat{\Phi}_{T,N,R,M}]
\ee
For instance, for a scalar field we have 
\ba \label{3.5}
W_{T,R}[F_{T,R},\Phi_{T,R}] &=& \exp(i \langle F_{T,R},\Phi_{T,R}\rangle)
\nonumber\\
\langle F_{T,R},\Phi_{T,R}\rangle&:=&\int_{\mathbb{R}}\; dt
\; \int_\sigma\; d^Dx\; F_{T,R}(t,x)\; \Phi(t,x)
\nonumber\\
& \Rightarrow &
\nonumber\\
W_{T,N,R,M}[\hat{F}_{T,N,R,M},\hat{\Phi}_{T,N,R,M}]
&=& \exp(i \langle \hat{F}_{T,N,R,M}, \hat{\Phi}_{T,N,R,M}\rangle),\;\;
\nonumber\\
\langle \hat{F}_{T,N,R,M}, \hat{\Phi}_{T,N,R,M}\rangle &:=&\epsilon_{R,M}^D\delta_{T,N} \sum_{n,m}
\hat{F}_{T,N,R,M}(n,m)\; \hat{\Phi}_{T,N,R,M}(n,m),\;\;
\nonumber\\
\hat{\Phi}_{T,N,R,M}(n,m) &:=& \langle \chi_{n\delta_{T,N},m\epsilon_{R,M}},\Phi\rangle
\ea
The discretised history field can now be used to define a probability
measure $\mu_{T,N,R,M}$ on the finite dimensional space 
$\Gamma_{T,N,R,M}$. In case of a scalar field we have obviously
$\Gamma_{T,N,R,M}=\mathbb{R}^{NM^D}$ and one will pick 
a measure on $\Gamma_{T,N,R,M}$ which is of the form:\\
(Lebesgue measure) times (damping factor) times (normalisation constant). 
The damping factor is usually motivated from the classical theory and 
is often the exponential of the negative of the discretised Euclidian
version of the action. Thus, to have something concrete in mind 
for a scalar field we consider a measure of the form 
\ba \label{3.6}
d\mu_{T,N,R,M}(\hat{\Phi}_{T,N,R,M}) &=&
\frac{\rho_{T,N,R,M}(\hat{\Phi}_{T,N,R,M})}{Z_{T,N,R,M}}\;
d^{NM^D}\hat{\Phi}_{T,N,R,M},\;\;
\nonumber\\
d^{NM^D}\hat{\Phi}_{T,N,R,M}
&:=&\prod_{n\in \mathbb{Z}_N,m\in \mathbb{Z}_M^D}
\; d\hat{\Phi}(n,m)
\ea
At this point we will not be specific about the choice of these three 
ingredients, but point out, that the discretisation of the Euclidian 
action, and thus the choice of $\mu_{T,N,R,M}$ is of course far from unique,
it is actually ambiguous. This is the case even if we impose that the formal
continuum limit of $\rho_{T,N,R,M}$ as $N,M\to \infty$ should be 
$\exp(-S_{T,R}[\Phi])$ where $S_{T,R}[\Phi]$ is the continuum 
Euclidian action restricted to $[0,T]\times\sigma_R$. In our concrete 
situation, such a discretisation could be chosen for instance as
\be \label{3.6a}
S_{T,N,R,M}[\hat{\Phi}_{T,N,R,M}]:=S[I_{T,N,R,M}\circ 
\frac{\hat{\Phi}_{T,N,R,M}}{\delta_{T,N} \epsilon^D_{R,M}}]
\ee
where the division by the cell volume reproduced the value 
$E_{T,N,R,M}\cdot \Phi$ in the continuum limit.
However, whatever ambiguous choice of discretisation one has made,
we insist that {\it the same} discretisation procedure has been applied 
{\it for all values of $T,N,R,M$}. For instance, if one has chosen 
to approximate a Laplacian by using only next neighbour field differences 
for one quadruple 
$(T_0,N_0,M_0,R_0)$ then one does it for all $(T,N,R,M)$
rather than arbitrarily taking also next to next neighbour differences into 
account for other values of $(T,N,R,M)$.
    
In this way we have produced a family of measures 
$\mu_{T,N,R,M}$ parametrised by $T,R,N,M$ where these numbers can 
all be arbitrarily large as long as they are finite. Each member 
of the family is specified by its corresponding generating 
functional
\be \label{3.7}
\mu_{T,N,R,M}(W_{T,N,R,M}[\hat{F}_{T,N,R,M}])
\ee
Now suppose that a continuum measure $\mu_{T,R}$ would exist
whose generating functional is $\mu_{T,R}(W_{T,R}[F_{T,R}])$ with 
$F_{T,R}\in
L_{T,R}$. In view of (\ref{3.4})
the injection map defines its {\it cylindrical projections}
\be \label{3.8}
\mu_{T,R}^{N,M}(W_{T,N,R,M}[\hat{F}_{T,N,R,M}])
:=\mu_{T,R}(W_{T,R}[I_{T,N,R,M}\cdot \hat{F}_{T,N,R,M}])
\ee
i.e. by restricting the smearing functions to be those of finite resolution.

This is the point where the coarse graining map comes into play. Namely 
we can combine injection and evaluation maps to produce the maps
\be \label{3.9}
I_{T,N\to N', R, M\to M'}:= 
E_{T,N', R, M'} \circ I_{T,N,R,M}: L_{T,N,R,M}\to L_{T,N',R,M'}
\ee
where $N,N',M,M'$ are arbitrary. Useful combinations of these numbers
are $N'=2^n N, M'= 2^m M,\; n,m\in \mathbb{N}_0$ because they correspond 
to viewing a function defined on the coarse lattice $(T,N,R,M)$ 
as a function on the finer lattice $(T,2^n N,R, 2^m M)$ of which the 
former is a sublattice. We will call them {\it coarse graining
maps} in this case. We will further impose the following 
restriction on the choice of the injection maps
\be \label{3.10}
I_{T,2^n N,R, 2^m M}\circ I_{T,N\to 2^n N,R,M\to 2^m M}=I_{T, N, R, M}
\ee
which is satisfied for the choice (\ref{3.2}). It means that 
considering a lattice function as a special continuum function 
is independent on what sublattice it is actually defined. 
To see this, we compute for $k\in \mathbb{Z}_{2^n N},\;
l\in \mathbb{Z}_{2^m M}^D$ 
\begin{align} \label{3.10a}
[I_{T,N\to 2^n N,R,M\to 2^m M}\cdot \hat{F}_{T,N,R,M}](k,l)
&=\sum_{k'\in \mathbb{Z}_N,l'\in \mathbb{Z}_M^D} \chi_{k' \delta_{T,N},
l'\epsilon_{R,M}}(k\delta_{T,2^n N},l\epsilon_{R,2^m M}) 
\hat{F}_{T, N, R, M}(k',l')
\nonumber\\ 
&\;= \hat{F}_{T, N, R, M}(\lfloor 2^{-n} k\rfloor,\lfloor 2^{-m} l\rfloor)
\end{align}
where the bracket is the Gauss bracket (separately for each argument), i.e.
$\lfloor 2^{-m} l\rfloor^a:=\lfloor 2^{-m} l^a\rfloor$. Accordingly 
\ba \label{3.10b}
&&(I_{T,2^n N,R, 2^m M}\circ I_{T,N\to 2^n N,R,M\to 2^m M}\cdot 
\hat{F}_{T, N, R, M})(t,x)
\nonumber\\
&=& 
\sum_{k\in \mathbb{Z}_{2^n N},l\in \mathbb{Z}_{2^m M}^D} 
\chi_{k \delta_{T, 2^n N},l \epsilon_{R, 2^m M}}(t,x)\;
\hat{F}_{T, N, R, M}(\lfloor 2^{-n} k \rfloor ,\lfloor 2^{-m} l\rfloor)
\nonumber\\
&=& 
\sum_{k'\in \mathbb{Z}_N,l'\in \mathbb{Z}_M^D}\;
\hat{F}_{T, N, R, M}(k',l')\;
\sum_{k\in \mathbb{Z}_{2^n N},l\in \mathbb{Z}_{2^m M}^D;\lfloor k 2^{-n}\rfloor =k';
\lfloor 2^{-m} l\rfloor =l'} 
\chi_{k \delta_{T, 2^n N},l \epsilon_{R, 2^m M}}(t,x)\;
\nonumber\\
&=& 
\sum_{k'\in \mathbb{Z}_N,l'\in \mathbb{Z}_M^D}\;
\hat{F}_{T, N, R, M}(k',l')\;
\sum_{k=k'+r, l=l'+s; r\in \{0,.., 2^n-1\};
s\in \{0,.., 2^m-1\}^D}
\chi_{k \delta_{T, 2^n N},l \epsilon_{R, 2^m M}}(t,x)\;
\nonumber\\
&=& 
\sum_{k'\in \mathbb{Z}_N,l'\in \mathbb{Z}_M^D}\;
\hat{F}_{T, N, R, M}(k',l')\;
\chi_{k' \delta_{T, N},l' \epsilon_{R,M}}(t,x)
\nonumber\\
&=& (I_{T,N,R,M}\cdot \hat{F}_{T,N,R,M})(t,x)
\ea 
as claimed.

As a consequence, we have the following {\it consistency condition}
among the coarse graining maps  
\ba \label{3.11}
&&I_{T, 2^n N\to 2^{n+n'} N,R, 2^m M\to 2^{m+m'} M} \circ  
I_{T, N\to 2^n N,R, M\to 2^m M} 
\nonumber\\
&=&
E_{T, 2^{n+n'} N, R, 2^{m+m'} M} \circ 
[I_{T, 2^n N, R, 2^m M}\circ I_{T,N\to 2^n N,R,M\to 2^m M}]
=E_{T, 2^{n+n'} N, R, 2^{m+m'} M} \circ I_{T, N, R, M}
\nonumber\\
&=& I_{T, N\to 2^{n+n'} N,R, M\to 2^{m+m'} M} 
\ea
It means that coarse graining can be done in steps over arbitrary intermediate
sublattices. 

For the cylindrical projections of $\mu_{T,R}$
this has the following consequence called {\it cylindrically consistency}
\ba \label{3.12}
&&\mu_{T,R}^{N,M}(W_{T,N,R,M}[F_{T,N,R,M}])
=\mu_{T,R}(W_{T,R}[I_{T,N,R,M}\cdot \hat{F}_{T,N,R,M}])
\nonumber\\
&=&\mu_{T,R}(W_{T,R}[I_{T,2^n N,R,2^m M}\cdot I_{T,N\to 2^n N, R, M\to 2^m M}
\circ \hat{F}_{T,N,R,M}])
\nonumber\\
&=&
\mu_{T,R}^{2^n N, 2^m M}(W_{T,2^n N,R,2^m M}[I_{T,N\to 2^n N, R, M\to 2^m M}
\cdot \hat{F}_{T,N,R,M}])
\ea
This means that a necessary condition for 
a {\it projective family} $(M,N)\mapsto 
(\mu_{T,R}^{M,N})$ to qualify as the cylindrical projections of a 
continuum measure, is that the family is cylindrically consistent. In suitable 
mathematical contexts, depending on the details of the theory under 
consideration, the condition is also sufficient \cite{Yam75}. Therefore,
cylindrical consistency is a very useful criterion in order to detect
the continuum limit of a projective family.

Renormalisation is now a construction principle to actually {\it derive} a 
cylindrically consistent family from a given family $\mu^{(0)}_{T,N,R,M}$, 
that was merely {\it defined} as above,  
by employing possibly natural but still ad hoc discretisation prescriptions.
We {\it construct} the sequence 
$(\mu^{(n)}_{T,N,R,M})_{n\in \mathbb{N}_0}$
inductively from $\mu^{(0)}_{T,N,R,M}$ by a {\it block spin transformation}
\be \label{3.13}
\mu^{(n+1)}_{T,N,R,M}(W_{T,N,R,M}[\hat{F}_{T,N,R,M}])
:=\mu^{(n)}_{T,2N,R,2M}(W_{T,2N,R,2M}[I_{T,N\to 2N,R, M\to 2M}
\cdot \hat{F}_{T,N,R,M}])
\ee
Note that (\ref{3.13}) indeed defines an entire new family of measures from 
the old one, because 1. each measure is completely defined in terms of its 
generating functional and 2. one performs (\ref{3.13}) coherently for all
$M,N$. The sequence of measures thus defined may or may not converge (here 
we use the notion of pointwise convergence on the $W_{T,N,R,M}[F_{T,N,R,M}]$).
If it or at least a subsequence converges, we call the limit a 
{\it fix point family} $\mu^\ast_{T,N,R,M}$.
The fix point family is cylindrically consistent in the sense
\ba \label{3.14}
&&\mu^\ast_{T,2N,R,2M}(W_{T,2N,R,2M}[I_{T,N\to 2N,R,M\to 2M}\cdot 
\hat{F}_{T,N,R,M}])
\nonumber\\
&=&\lim_{n\to\infty}
\mu^{(n)}_{T,2N,R,2M}(W_{T,2N,R,2M}[I_{T,N\to 2N,R,M\to 2M}\cdot 
\hat{F}_{T,N,R,M}])
\nonumber\\
&=&\lim_{n\to\infty}
\mu^{(n+1)}_{T,N,R,M}(W_{T,N,R,M}[\hat{F}_{T,N,R,M}])
=\mu^\ast_{T,N,R,M}(W_{T,N,R,M}[\hat{F}_{T,N,R,M}])
\ea
and thus defines a continuum measure $\mu^\ast_{T,R}$ under the afore 
mentioned conditions.

Note that the fix point family of course depends on the parameters 
$N,M$. The measure $\mu^\ast_{T,N,R,M}$ at given $N,M$ {\it is} the 
continuum measure, but restricted to observables, that is, field probes 
$F_{T,N,R,M}=I_{T,N,R,M}\cdot \hat{F}_{T,N,R,M}$, of finite resolution. 
In the physics jargon customary in renormalisation, these are the so-called 
{\it perfect measures} (or {\it perfect actions} if one prefers to think
in terms of actions rather than measures). The terminology intends 
to express the fact that although one measures at finite resolution, the 
measurements are still described by the continuum theory \cite{Has98,Has08,HN93} .

Finally, we comment on the notion of universality. The first issue concerns 
the dependence of the fix point structure of the theory on the choice of 
coarse graining map. The coarse graining maps that we took into account 
all were based on injections of finite resolution smearing functions into
the continuum, subject to the conditions (\ref{3.2a}) and (\ref{3.10}) and 
thus are of a particular class. One would like to know, how much 
the fixed point structure depends on the particulars of the choice within this class.
The second issue concerns the dependence of the fixed point structure on the 
initial family $\mu^{(0)}_{T,N,R,M}$. 
In \cite{21,22} we study both dependencies for the case of the Klein-Gordon field.
To finish this section, let us mention that what has been said here 
can be generalised. Instead of labelling finite resolution smearing functions
by integers $M,N\in \mathbb{N}_0^2$ 
we could use any partially ordered 
set $\cal I$ which can be partitioned into directed subsets. 
Here partially ordered means that there is a relation
(symmetric, reflexive, transitive)
$i<j$ between certain but not necessarily all elements $i,j\in {\cal I}$    
and a subset ${\cal J}\subset {\cal I}$ is said to be 
directed if for any $i,j \in {\cal J}$ there exists 
$k\in {\cal J}$ such that $i,j<k$. In our case we say that $(M,N)<(M',N')$
if $M'=2^m M,N'= 2^n N$ for some $(m,n)\in \mathbb{N}_0^2$ and thus the 
directed subsets of ${\cal I}$ are of the form 
${\cal J}_{M,N}=\{(2^m M, 2^n N),\; m,n\in \mathbb{N}_0^2\}$ where 
$M,N$ are odd, positive integers. Evidently these sets are mutually disjoint
and partition $\cal I$.

Dropping the labels $R,T$ for convenience
we would then have for function spaces $L, L_j$ 
corresponding injections $I_j$, evaluations $E_j$,
coarse grainings $I_{ij}=E_j \circ I_i$ for $i<j$ in the same directed subset
subject to $I_{ii}={\rm id}_{L_i}$ and $I_j \circ I_{ij}=I_i$.    
From here on, everything stays the same. Such a more abstract viewpoint 
is convenient, if one would like to define more general coarse graining 
transformations.

\section{Hamiltonian renormalisation}
\label{s4}
 
In this section we intend to 
define Hamiltonian renormalisation in such a way, that 
the following diagram closes: Given a family of 
discretisations of the OS Data
$({\cal H}^{(0)}_{R,M}, \Omega^{(0)}_{R,M}, H^{(0)}_{R,M})$ labelled by $M$ 
one can construct the corresponding Wiener measure family $\mu^{(0)}_{R,M}$.
When restricted to smearing functions of time support $T$ and finite 
time resolution labelled by $N$ one obtains its temporal cylindrical 
projections $\mu_{T,N,R,M}$ family. We notice that while we can make 
contact to the previous section, the discretised measures that we obtain are of 
a special type: They are already cylindrically consistent with respect
to coarse grainings in the time direction. The special role of time cannot
be avoided in the Hamiltonian setting because the Hamiltonian generator 
of a contraction semi-group needs a continuous parameter and cannot be derived
using a temporal lattice. Accordingly, we drop the label $T$ for the rest of 
this section assuming that the purely temporal renormalisation was 
already carried out.

One can now perform path integral renormalisation as outlined in the previous 
section and obtain a fix point measure $\mu_{R}$. We will see that this 
measure, if it exists, is automatically reflection positive if the initial
family is. From it we then obtain the continuum 
OS data $({\cal H}_R, \Omega_R, H_R)$ by OS reconstruction.
What we intend to find out is how to pass from the starting point
$({\cal H}^{(0)}_{R,M}, \Omega^{(0)}_{R,M}, H^{(0)}_{R,M})$ directly to the 
same endpoint 
to $({\cal H}_R, \Omega_R, H_R)$. In other words, the direct 
renormalisation from
$({\cal H}^{(0)}_{R,M}, \Omega^{(0)}_{R,M}, H^{(0)}_{R,M})$ to 
$({\cal H}_R, \Omega_R, H_R)$ allows us to define the corresponding
Wiener measure $\mu_R$ and that measure coincides 
with the one that one obtains via path integral renormalisation.

The advantage of such a scheme, if it exists, is quite obvious:
The construction of the complicated 
Wiener measures can be avoided altogether and 
one can stay entirely within the Hamiltonian setting. The disadvantage 
maybe is, that three ingredients must be renormalised instead of 
only one: The Hilbert space, 
the vacuum and the Hamiltonian annihilating that vacuum. By contrast,
path integral renormalisation only needs to renormalise the measure.

Unfortunately, we will only be partially successful in doing so: 
While we are able to find an exact Hamiltonian renormalisation scheme 
that is equivalent to the path integral renormalisation scheme,
this scheme still makes non-trivial reference to the Wiener measures at
each renormalisation step. That is, while the Wiener measure is a derived concept 
fully determined by OS data, one still needs to construct it 
in an intermediate step. However, the derived renormalisation scheme 
motivates a {\it different scheme} which {\it does} stay entirely 
within the Hamiltonian framework without the need to construct the 
Wiener measure. Of course, there is no guarantee that the two 
schemes have the same fixed points. Nevertheless, for the second scheme,
if it does have fixed points, these do define again Wiener measures 
in the continuum satisfying reflection positivity. Even for free field 
theories
the renormalisation trajectories of both schemes are different, however
their fixed points turn out to be identical. \\
\\
The general setting is as follows:\\
\\
From a given family of OS data $({\cal H}^{(0)}_{R,M}, 
\Omega^{(0)}_{R,M}, H^{(0)}_{R,M})$ let us construct the Wiener measure 
family $\mu^{(0)}_{R,M}$. That is,
for a smearing function $F_{R,M}(\beta)=\sum_{k=1}^n f^k_{R,M} 
\delta(\beta,\beta_k)$ of discrete and finite 
time support $\beta_1<..<\beta_n$ and finite spatial resolution
\be \label{4.1}
\mu^{(0)}_{R,M}(W_{R,M}[\hat{F}_{R,M}])=
\langle \Omega^{(0)}_{R,M}, w_{R,M}[f^n_{R,M}] 
e^{-(\beta_n-\beta_{n-1})H^{(0)}_{R,M}}
... e^{-(\beta_2-\beta_1)H^{(0)}_{R,M}} w_{R,M}[f^1_{R,M}]
\Omega^{(0)}_{R,M}\rangle_{{\cal H}^{(0)}_{R,M}}
\ee
As the measure is already temporally cylindrically consistent, 
we restrict the injection and evaluation maps of the previous section to 
the spatial variables. Thus, we have function spaces $L_R, L_{R,M}$ and the 
maps 
\be \label{4.2}
I_{R,M}:\; L_{R,M}\to L_R;\; E_{R,M}:\; L_R\to L_{R,M}
\ee
subject to the constraint
\be \label{4.3}
E_{R,M}\circ I_{R,M}={\rm id}_{L_{R,M}}
\ee
from which one obtains the coarse graining maps
\be \label{4.4}
I_{R,M\to 2^m M}=E_{R, 2^m M}\circ I_{R,M}
\ee
subject to the constraints
\be \label{4.5}
I_{R,2^m M}\circ I_{R,M\to 2^m M}=I_{R,M}
\ee
These maps are related to the ones for path integral renormalisation 
including the time direction 
by the assumption that the latter factorises into maps for the time direction
and the above maps for the spatial direction, see (\ref{3.1}). Then 
one carries out first only temporal renormalisation and arrives at 
the purely temporal fixed point $\mu^{(0)}_{R,M}$ for each spatial cut-off $M$
which are equivalent to the above OS data.\\  
\\
In the next subsection we derive first the Hamiltonian renormalisation scheme 
which is indeed equivalent to the path integral renormalisation scheme. We call 
it the ``Path Integral Induced Hamiltonian Renormalisation Scheme''.
This motivates the second subsection in which we define another 
renormalisation scheme which we coin ``Direct Hamiltonian Renormalisation
Scheme".

\subsection{Derivation of the Path Integral Induced Hamiltonian Renormalisation 
Scheme}
\label{s4.1}

We begin by showing that the path integral renormalisation based on the 
coarse graining maps $I_{R, M\to 2M}$ does not leave the space of 
families of time translation invariant, reflection invariant, reflection 
positive and reflection continuous 
measures if the initial family $\{\mu^{(0)}_{R,M}\}_{M\in \mathbb{N}}$
is of this kind.

Recall that path integral renormalisation is defined by a sequence 
$n\mapsto \mu^{(n)}_{R,M}$ of families $\{\mu^{(n)}_{R,M}\}_{M\in \mathbb{N}}$
defined recursively by 
\be \label{4.100}
\mu^{(n+1)}_{R,M}(W_{R,M}[F_{R,M}]):=\mu^{(n)}_{R,2M}(
W_{R,2M}(I_{R,M\to 2M}\cdot F_{R,M}])
\ee
where $I_{R,M\to 2M}$ acts simultaneously at each time step only 
on the spatial arguments of $F_{R,M}$ and thus 
commutes with time reflection $\theta$ and time translation $T_s$.  \\
Time-Translation-Invariance:
\ba \label{4.101}
\mu^{(n+1)}_{R,M}({\cal U}(s)\cdot W_{R,M}[F_{R,M}])
&=& 
\mu^{(n+1)}_{R,M}(W_{R,M}[T_s\cdot F_{R,M}])
\nonumber\\
&=& \mu^{(n)}_{R,2M}(W_{R,2M}[I_{R,M\to 2M}\cdot T_s\cdot F_{R,M}])
\nonumber\\
&=& \mu^{(n)}_{R,2M}(W_{R,2M}[T_s\cdot I_{R,M\to 2M}\cdot F_{R,M}])
\nonumber\\
&=& \mu^{(n)}_{R,2M}({\cal U}(s)\cdot W_{R,2M}[I_{R,M\to 2M}\cdot F_{R,M}])
\nonumber\\
&=& \mu^{(n)}_{R,2M}(W_{R,2M}[I_{R,M\to 2M}\cdot F_{R,M}])
\nonumber\\
&=&
\mu^{(n+1)}_{R,M}(W_{R,M}[F_{R,M}])
\ea
Time-Reflection-Invariance:
\ba \label{4.102}
\mu^{(n+1)}_{R,M}(R\cdot W_{R,M}[F_{R,M}])
&=& 
\mu^{(n+1)}_{R,M}(W_{R,M}[\theta\cdot F_{R,M}])
\nonumber\\
&=& \mu^{(n)}_{R,2M}(W_{R,2M}[I_{R,M\to 2M}\cdot \theta\cdot F_{R,M}])
\nonumber\\
&=& \mu^{(n)}_{R,2M}(W_{R,2M}[\theta\cdot I_{R,M\to 2M}\cdot F_{R,M}])
\nonumber\\
&=& \mu^{(n)}_{R,2M}(R\cdot W_{R,2M}[I_{R,M\to 2M}\cdot F_{R,M}])
\nonumber\\
&=& \mu^{(n)}_{R,2M}(W_{R,2M}[I_{R,M\to 2M}\cdot F_{R,M}])
\nonumber\\
&=&
\mu^{(n+1)}_{R,M}(W_{R,M}[F_{R,M}])
\ea
Reflection Positivity:
\ba \label{4.103}
&&\mu^{(n+1)}_{R,M}(W_{R,M}[F_{R,M}]^\ast \; (R\cdot W_{R,M}[F_{R,M}]))
= \sum_\alpha z_\alpha 
\mu^{(n+1)}_{R,M}(W_{R,M}[F_{\alpha,R}(F_{R,M}^\ast, \theta\cdot F_{R,M})])
\nonumber\\
&=& \sum_\alpha z_\alpha 
\mu^{(n)}_{R,2M}(W_{R,2M}[I_{R,M\to 2M}\cdot 
F_{\alpha,R}(F_{R,M}^\ast, \theta\cdot F_{R,M})])
\nonumber\\
&=& \sum_\alpha z_\alpha 
\mu^{(n)}_{R,2M}(W_{R,2M}[F_{\alpha,R}(
(I_{R,M\to 2M}\cdot F_{R,M})^\ast, \theta\cdot I_{R,M\to 2M}\cdot F_{R,M})])
\nonumber\\
&=&  
\mu^{(n)}_{R,2M}((W_{R,2M}[I_{R,M\to 2M}\cdot F_{R,M}])^\ast \; 
(R\cdot W_{R,M}[I_{R,M\to 2M}\cdot F_{R,M}]))\nonumber
\\
&\ge & 0
\ea
Here we actually treated only a special case, namely that we just have 
a single function $F_{R,M}$ with a single time dependence. However, the 
general case is completely analogous and just requires the Weyl element 
decomposition at each coinciding point of time and will be left to the 
reader. We have assumed the equivariance of the Weyl algebra relations 
(\ref{2.19}) and thus its spacetime equivalent with respect to the 
coarse graining maps, that is
\be \label{4.103a}
I_{R,M\to 2M}\cdot F^\ast_{R,M}=
(I_{R,M\to 2M}\cdot F_{R,M})^\ast,\;
I_{R,M\to 2M}\cdot F_{\alpha,R}(F_{R,M},F'_{R,M})
=F_{\alpha,R}(I_{R,M\to 2M}\cdot F_{R,M},I_{R,M\to 2M}\cdot F'_{R,M})
\ee
which is satisfied in all known examples. 
Note that the Weyl relations (\ref{2.19}) of the continuum induce the same 
relations on the finite resolution analogues due to 
\be \label{4.103b}
W_R[I_{R,M}\cdot F_{R,M}]=:W_{R,M}[F_{R,M}]
\ee
in particular the constants $z_\alpha$ and the composition functions 
$F_\alpha$ do not depend on $M$. That reflection continuity (or equivalently 
time translation continuity) is also 
inherited is obvious since again time translations commute with the coarse 
graining map and thus the corresponding semigroups have a generator.

It follows that each of the measures $\mu^{(n)}_{R,M}$ defines OS
data $({\cal H}^{(n)}_{R,M}, \Omega^{(n)}_{R,M}, H^{n)}_{R,M})$ by 
OS reconstruction and we now want to describe them in more detail.

Let $[W_{R,M}[F_{R,M}]]_{\mu^{(n)}_{R,M}}\in {\cal H}^{(n)}_{R,M}$ denote 
the equivalence class 
of the vector $W_{R,M}[F_{R,M}]$, T-supp$(F_{R,M})\subset \mathbb{R}_+$
with respect to the reflection positive inner product defined by 
$\mu^{(n)}_{R,M}$, in particular $\Omega^{(n)}_{R,M}=[1]_{\mu^{(n)}_{R,M}}$. 
Let 
\be \label{4.104}
J^{(n)}_{R,M\to 2M}:\; 
{\cal H}^{(n+1)}_{R,M}
\to {\cal H}^{(n)}_{R,2M};\;\;
[W_{R,M}[F_{R,M}]]_{\mu^{(n+1)}_{R,M}}\mapsto
[W_{R,2M}[I_{R,M\to 2M}\cdot F_{R,M}]]_{\mu^{(n)}_{R,2M}}
\ee
This defines a sequence of densely defined families of 
embeddings of Hilbert spaces. Then by the definition of the 
reflection positive inner product and the assumed Weyl algebra relations 
by a similar calculation as in (\ref{4.103})
\ba \label{4.105}
&&\langle [W_{R,M}[F_{R,M}]]_{\mu^{(n+1)}_{R,M}},\; 
[W_{R,M}[F'_{R,M}]]_{\mu^{(n+1)}_{R,M}}\rangle_{{\cal H}^{(n+1)}_{R,M}} 
= \mu^{(n+1)}_{R,M}((W_{R,M}[F_{R,M}])^\ast W_{R,M}[\theta\cdot F'_{R,M}])
\nonumber\\
&=& \mu^{(n)}_{R,2M}((W_{R,2M}[I_{R,M\to 2M}\cdot F_{R,M}])^\ast 
W_{R,2M}[\theta\cdot I_{R,M\to 2M}\cdot  F'_{R,M}])
\nonumber\\
&=&
\langle [W_{R,2M}[I_{R,M\to 2M}\cdot F_{R,M}]]_{\mu^{(n)}_{R,2M}},\; 
[W_{R,2M}[I_{R,M\to 2M}\cdot F'_{R,M}]]_{\mu^{(n)}_{R,2M}}\rangle_{{\cal H}^{(n)}_{R,2M}} 
\nonumber\\
&=& \langle J^{(n)}_{R,M} [W_{R,M}[F_{R,M}]]_{\mu^{(n+1)}_{R,M}},\; 
J^{(n)}_{R,M} [W_{R,M}[F'_{R,M}]]_{\mu^{(n+1)}_{R,M}}\rangle_{{\cal H}^{(n)}_{R,2M}} 
\ea
It follows that for each $n$ the family of maps $J^{(n)}_{R,M\to 2M}$ 
defines a family of isometric embeddings ${\cal H}^{(n+1)}_{R,M}\to
{\cal H}^{(n)}_{R,2M}$, i.e. 
\be \label{4.106}
(J^{(n)}_{R,M\to 2M})^\dagger J^{(n)}_{R,M\to 2M}=1_{{\cal H}^{(n+1)}_{R,M}}
\ee
It follows that 
\be \label{4.107}
P^{(n)}_{R,M\to 2M}:=J^{(n)}_{R,M\to 2M}\; (J^{(n)}_{R,M\to 2M})^\dagger
\ee
are projections onto the subspace 
$J^{(n)}_{R,M\to 2M} {\cal H}^{(n+1)}_{R,M}$ of 
${\cal H}^{(n)}_{R,2M}$, that is
\be \label{4.108}
(P^{(n)}_{R,M\to 2M})^2=
J^{(n)}_{R,M\to 2M}\; \{(J^{(n)}_{R,M\to 2M})^\dagger\;
J^{(n)}_{R,M\to 2M}\}\; (J^{(n)}_{R,M\to 2M})^\dagger
=P^{(n)}_{R,M\to 2M}
\ee
For later reference we also note the identities 
\be \label{4.108a}
P^{(n)}_{R,M\to 2M} \; J^{(n)}_{R,M\to 2M}=J^{(n)}_{R,M\to 2M}\;\;
\Rightarrow\;\;
(J^{(n)}_{R,M\to 2M})^\dagger\; P^{(n)}_{R,M\to 2M}=
(J^{(n)}_{R,M\to 2M})^\dagger
\ee

Next we find for the flow of the Hamiltonians by the already familiar 
calculations
\ba \label{4.109}
&&
\langle [W_{R,M}[F_{R,M}]]_{\mu^{(n+1)}_{R,M}},\; e^{-\beta H^{(n+1)}_{R,M}}\;
[W_{R,M}[F'_{R,M}]]_{\mu^{(n+1)}_{R,M}}\rangle_{{\cal H}^{(n+1)}_{R,M}} 
\nonumber\\
&=&
\mu^{(n+1)}_{R,M}((W_{R,M}[F_{R,M}])^\ast \;
W_{R,M}[\theta\cdot T_\beta\cdot 
F'_{R,M}])
\nonumber\\
&=&
\mu^{(n)}_{R,2M}((W_{R,2M}[I_{R,M\to 2M}\cdot F_{R,M}])^\ast\;
W_{R,2M}[I_{R,M\to 2M}\cdot\theta\cdot T_\beta\cdot 
F'_{R,M}])
\nonumber\\
&=&
\mu^{(n)}_{R,2M}((W_{R,2M}[I_{R,M\to 2M}\cdot F_{R,M}])^\ast\;
W_{R,2M}[\theta\cdot T_\beta\cdot I_{R,M\to 2M}\cdot F'_{R,M}])
\nonumber\\
&=&
\langle [W_{R,2M}[I_{R,M\to 2M}\cdot F_{R,M}]]_{\mu^{(n)}_{R,2M}},\;
e^{-\beta H^{(n)}_{R,2M}}\; [W_{R,2M}[I_{R,M\to 2M}\cdot 
F'_{R,M}]]_{\mu^{(n)}_{R,2M}}\rangle_{{\cal H}^{(n)}_{R,2M}} 
\nonumber\\
&=&
\langle J^{(n)}_{R,M\to 2M}\;[W_{R,M}[F_{R,M}]]_{\mu^{(n+1)}_{R,2M}},\;
e^{-\beta H^{(n)}_{R,2M}}\; J^{(n)}_{R,M\to 2M}\; [W_{R,M}[ 
F'_{R,M}]]_{\mu^{(n+1)}_{R,M}}\rangle_{{\cal H}^{(n)}_{R,2M}} 
\ea
whence for all $\beta>0$
\be \label{4.110}
e^{-\beta H^{(n+1)}_{R,M}}=(J^{(n)}_{R,M\to 2M})^\dagger\;
e^{-\beta H^{(n)}_{R,2M}}\;J^{(n)}_{R,M\to 2M}
\ee
while for $\beta=0$ we recover the isometry condition (\ref{4.106}).

Taking the $l$-th derivative\footnote{This makes sense on analytic 
vectors that exist owing to the granted self-adjointness of the 
Hamiltonians.
If one wants to avoid domain questions, then one could carry out the same 
analysis that follows from here on and would conclude that 
$[P^{(n)}_{R,M\to 2M},e^{-\beta H^{(n)}_{R,2M}}]=0$ for all $\beta$. 
Thus all spectral projections of $H^{(n)}_{R,2M}$ commute with 
$P^{(n)}_{R,2M}$ which is exactly what is meant when talking about the 
commutativity of unbounded operators.} of (\ref{4.110}) at $\beta=0$ (which 
is possible on a dense and invariant domain of the Hamiltonian 
(analytic vectors) due to the preservation of the OS continuity 
during renormalisation) we find 
\be \label{4.111}
(H^{(n+1)}_{R,M})^l=
(J^{(n)}_{R,M\to 2M})^\dagger\;
(H^{(n)}_{R,2M})^l\;J^{(n)}_{R,M\to 2M}
\ee
For $l=1$ we conclude that the sequence of families of Hamiltonians 
remains symmetric if the initial family is.
Combining (\ref{4.111}) for $l=1,2$ we find 
\be \label{4.112}
(J^{(n)}_{R,M\to 2M})^\dagger\;
(H^{(n)}_{R,2M})^2\;J^{(n)}_{R,M\to 2M}
=(J^{(n)}_{R,M\to 2M})^\dagger\;
H^{(n)}_{R,2M}\; \;P^{(n)}_{R,M\to 2M}\;
H^{(n)}_{R,2M}\; \;J^{(n)}_{R,M\to 2M}
\ee
Multiplying (\ref{4.112}) with $(J^{(n)}_{R,M\to 2M})^\dagger,
J^{(n)}_{R,M\to 2M}$ from the right and left respectively yields the 
identity
\be \label{4.112a}
P^{(n)}_{R,M\to 2M}\;
(H^{(n)}_{R,2M})^2\;P^{(n)}_{R,M\to 2M}
=P^{(n)}_{R,M\to 2M}\;
H^{(n)}_{R,2M}\; \;P^{(n)}_{R,M\to 2M}\;
H^{(n)}_{R,2M}\; \;P^{(n)}_{R,M\to 2M}
\ee
which is equivalent to 
\ba \label{4.113}
0 &=&
P^{(n)}_{R,M\to 2M}\;
H^{(n)}_{R,2M}\; [1_{{\cal H}^{(n)}_{R,2M}}-P^{(n)}_{R,M\to 2M}]\;
H^{(n)}_{R,2M}\; \;P^{(n)}_{R,M\to 2M}
\\
&=&
P^{(n)}_{R,M\to 2M}\;
H^{(n)}_{R,2M}\; [P^{(n)}_{R,M\to 2M}]_\perp\;
H^{(n)}_{R,2M}\; \;P^{(n)}_{R,M\to 2M}
\nonumber\\
&=&
(P^{(n)}_{R,M\to 2M}\;
H^{(n)}_{R,2M}\; [P^{(n)}_{R,M\to 2M}]_\perp)\;
(P^{(n)}_{R,M\to 2M}\;
H^{(n)}_{R,2M}\; [P^{(n)}_{R,M\to 2M}]_\perp)^\dagger=:A^\dagger\; A
\nonumber
\ea
It follows $\langle \Psi_{R,2M},A^\dagger A \Psi_{R,2M}\rangle=||A\Psi_{R,2M}||^2=0$
for all $\Psi_{R,2M}$ in the domain of $A$ so that $A=0$. 
It follows 
\be \label{4.114}
H^{(n)}_{R,2M} P^{(n)}_{R,M\to 2M}=
P^{(n)}_{R,M\to 2M} H^{(n)}_{R,2M} P^{(n)}_{R,M\to 2M}
=(P^{(n)}_{R,M\to 2M} H^{(n)}_{R,2M} P^{(n)}_{R,M\to 2M})^\dagger
=P^{(n)}_{R,M\to 2M} \; H^{(n)}_{R,2M}
\ee
so that $H^{(n)}_{R,2M}$ has the remarkable property to 
preserve the subspace $P^{(n)}_{R,M\to 2M} {\cal H}^{(n)}_{R,2M}$. This 
explains why (\ref{4.110}) is consistent with the semigroup property.
We note that from the semigroup property
\begin{align} \label{4.115}
e^{-(\beta_1+\beta_2)H^{(n+1)}_{R,M}}&=\;
(J^{(n)}_{R,M\to 2M})^\dagger\;\;
e^{-(\beta_1+\beta_2)H^{(n)}_{R,2M}}\;
J^{(n)}_{R,M\to 2M}
\nonumber\\
=\; e^{-\beta_1 H^{(n+1)}_{R,M}}\;
e^{-\beta_2 H^{(n+1)}_{R,M}}
&=\;(J^{(n)}_{R,M\to 2M})^\dagger\;\;
e^{-\beta_1 H^{(n)}_{R,2M}}\; P^{(n)}_{R,M\to 2M}
e^{-\beta_2 H^{(n)}_{R,2M}}\; J^{(n)}_{R,M\to 2M}
\end{align}

This property is at first sight indeed astonishing and in fact counter
intuitive: Suppose that for all $n\in \mathbb{N}$ a representative 
of $[W_{R,M}[F_{R,M}]]_{\mu^{(n)}_{R,M}}$ is given by a linear 
combination of vectors of the form
$w_{R,M}[f_{R,M}]\Omega^{(n)}_{R,M}$ for suitable $f_{R,M}$ with say 
$\Omega^{(n)}_{R,M}=1,\;\forall M,n$ in the ground state representation, i.e. 
in terms of sharp time zero fields on a lattice labelled by $M$. Then   
$J^{(n)}_{R,M\to 2M}$ is actually independent of $n$ and 
basically embeds those vectors into the lattice labelled
by $2M$ using the map $I_{R,M\to 2M}$ e.g. by associating a constant value 
to the smearing fields in a block within the fine lattice that corresponds to
a point in the coarse lattice. Then (\ref{4.114}) would imply that 
the Hamiltonian on the fine lattice actually preserves the space of 
vectors with smearing functions of that special type and that property 
is certainly violated in all applications.

The apparent contradiction is resolved by noticing that there is 
no guarantee that $[W_{R,M}[F_{R,M}]]_{\mu^{(n)}_{R,M}}$ can in fact
be represented by vectors in the linear span of the 
$w_{R,M}[f_{R,M}]\Omega^{(n)}_{R,M}$. This is because of the innocent 
looking dependence of the equivalence class $[.]_{\mu^{(n)}_{R,M}}$ on 
the measure. Even if the initial equivalence class can be represented in 
terms of sharp time zero fields of a single field species, this may no longer 
be the case after already one renormalisation step. Indeed, this is what
happens already in the free field theory case. There one can label  
$[W_{R,M}[F_{R,M}]]_{\mu^{(n)}_{R,M}}$ either as a single field species 
at an increasing number of sharp times (as $n\to \infty$)
or as a single field
at single sharp time zero which is composed of an increasing number of 
mutually interacting free field species. Thus 
the intuition that one may have with the meaning of $J^{(n)}_{R,M\to 2M}$
is quite misleading.

For the same reason, the quite compactly looking condition (\ref{4.111}) 
for $l=1$ only partly fulfils our aim to directly compute the flow 
of the Hamiltonians purely within the Hamiltonian theory:\\
Starting with initial OS data 
$({\cal H}^{(0)}_{R,M},\Omega^{(0)}_{R,M},H^{(0)}_{R,M})$ we have to 
compute the corresponding Wiener measures $\mu^{(0)}_{R,M}$ in order 
to define $J^{(0)}_{R,M\to 2M}$ via (\ref{4.104}) As alluded to above, this 
may not only drastically change the meaning of $w_{R,M}[f_{R,M}]$ if we want
to keep a sharp zero-time interpretation of the equivalence classes, 
we also cannot avoid the computation of the Wiener measure as we 
actually intended to do.   

We close this subsection with the remark that the path integral induced 
Hamiltonian renormalisation discussed above bears a certain similarity with the 
Hamiltonian renormalisation scheme advocated in \cite{GlaWil,Wegner} 
in the sense that both schemes attempt to push the renormalisation 
flow of the Hamiltonian into a block diagonal form. The difference is that,
in those latter schemes in contrast to what happens here, first this is done 
for two fixed resolutions (the 
high energy cut-off and the low energy effective resolution) and not for 
a whole family thereof and second without changing the ``size''and the 
vacuum of the Hilbert space at the same resolution. A better understanding 
of the relation between these two schemes and the one we will propose in the 
next section is desirable and will be subject of a future publication.

\subsection{Motivation of the Direct Hamiltonian Renormalisation Scheme}
\label{s4.2}

The caveat to avoiding the computation of the Wiener measure pointed out 
in the previous subsection and the complications associated with the 
changing realisation of the equivalence classes of vectors under 
the path integral renormalisation flow suggest a natural more direct
Hamiltonian renormalisation scheme which stays maximally close 
to the path integral renormalisation scheme:\\ 
\\
Starting from initial OS data $({\cal H}^{(0)}_{R,M},\;\Omega^{(0)}_{R,M},\;
H^{(0)}_{R,M})$ we define the maps
\be \label{4.116}
j^{(n)}_{R,M\to 2M}:\;{\cal H}^{(n+1)}_{R,M}\to {\cal H}^{(n)}_{R,2M};\;
w_{R,M}[f_{R,M}]\Omega^{(n+1)}_{R,M}\mapsto   
w_{R,2M}[I_{R,M\to 2M}\cdot f_{R,M}]\Omega^{(n)}_{R,2M}
\ee
in terms of Weyl elements of 
single species sharp time zero fields which are implicitly 
assumed to span ${\cal H}^{(n)}_{R,M}$ for all $n$. In particular for 
$f_{R,M}=0$ 
\be \label{4.116a}
j^{(n)}_{R,M\to 2M} \;\Omega^{(n+1)}_{R,M}=\Omega^{(n)}_{R,2M}
\ee
Definition 
(\ref{4.116}) replaces definition (\ref{4.104}). Note that the actual
action of (\ref{4.116}), in contrast to $J^{(n)}_{R,M}$ 
does not depend on $n$ and is purely geometrically defined entirely
in terms of $I_{R,M\to 2M}$, except for (\ref{4.116a}). In the ground state
representation, however, $\Omega^{(n)}_{R,M}=1$ for all $n,M$.
  
Next we insist on the analogue of the isometry condition (\ref{4.106})
\be \label{4.117}
(j^{(n)}_{R,M\to 2M})^\dagger\; j^{(n)}_{R,M\to 2M}=1_{{\cal H}^{(n+1)}_{R,M}}
\ee
which in practical terms leads to a flow 
of Hilbert space measures (recall that \\
$\nu_{R,M}(.)=\langle\Omega_{R,M},\;.\;
\Omega_{R,M}\rangle_{{\cal H}_{R,M}}$)
\be \label{4.118a}
\nu^{(n+1)}_{R,M}(w_{R,M}[f_{R,M}])=
\nu^{(n)}_{R,2M}(w_{R,2M}[I_{R,M\to 2M}\cdot f_{R,M}])
\ee
Finally, we impose the analogue of (\ref{4.111}) for $l=1$ 
\be \label{4.118}
H^{(n+1)}_{R,M}:=(j^{(n)}_{R,M\to 2M})^\dagger\; H^{(n)}_{R,M}\; 
j^{(n)}_{R,M\to 2M}
\ee
but not the analogue of the stronger condition (\ref{4.110}) since it would 
imply $[H^{(n)}_{R,2M},p^{(n)}_{R,M\to 2M}]=0$ for the projector 
$p^{(n)}_{R,M\to 2M}=j^{(n)}_{R,M\to 2M} (j^{(n)}_{R,M\to 2M})^\dagger$ which 
is not physically viable as we have argued above.

Note that the flow of Hamiltonians and vacua is consistent since 
\be \label{4.119}
H^{(n+1)}_{R,M} \Omega^{(n+1)}_{R,M}=
(j^{(n)}_{R,M\to 2M})^\dagger\; H^{(n)}_{R,2M} 
\;[j^{(n)}_{R,M\to 2M} \Omega^{(n+1)}_{R,M}]
=(j^{(n)}_{R,M\to 2M})^\dagger\; H^{(n)}_{R,2M} 
\;\Omega^{(n)}_{R,2M}=0
\ee

Conditions (\ref{4.116}), (\ref{4.117}) and (\ref{4.118}) define a flow 
of OS data or equivalently Wiener measures which is in close analogy to the
conditions (\ref{4.106}), (\ref{4.104}) and (\ref{4.111}) stated in the 
previous subsection. However, that flow will generically not 
coincide with the flow defined in the previous section, already 
not for the free field. However, we are not interested in the renormalisation 
trajectory but rather in its fixed points, i.e. we want to know whether
the direct and the path integral Hamiltonian flow lead to the same fixed 
points. At least for the free field theory case the answer is in the 
affirmative and the way this happens is quite interesting: Namely,
while the direct Hamiltonian flow always stays within a single field 
species context, the path integral flow develops an infinite number 
of field species with specific coupling ``constants'', the non-vanishing 
number of which increase with each renormalisation step. These constants 
play a role similar to the ones of the K\"allen-Lehmann spectral function
of a generalised free field. However, all
but a single member of these fields develop an infinite mass in the 
continuum limit $M\to \infty$ and thus decouple. The remaining single 
field of finite mass of the path integral flow can be identified with the 
one of the 
direct Hamiltonian flow in the continuum limit. For this reason, both schemes 
result in the same continuum theory which of course coincides with the 
free field of a single finite mass. 

What can be said about the fixed points of the direct Hamiltonian flow
when they exist?
Let us denote the fixed point structures by $j_{R, M\to 2M},\;
{\cal H}_{R,M},\;\Omega_{R,M},\;H_{R,M}$. Then 
\be \label{4.120}
j_{R,M\to 2^n M}:=j_{R,2^{n-1} M\to 2^n M}\cdot..\cdot j_{R,M\to 2M}:\; 
{\cal H}_{R,M}\to {\cal H}_{R,2^n M}
\ee
defines a family of 
isometric injections of Hilbert spaces labelled by the 
partially ordered set $M:=(k,n):=(2k+1)\; 2^n,\; k,n\in 
\mathbb{N}_0$ where $M<M'$ iff $k=k'$ and $n<n'$. It has directed (in fact 
linearly ordered) subsets, namely those for fixed $k$.
The injections satisfy the 
consistency condition 
\be \label{4.121}
j_{R,M_2\to M_3}\cdot j_{R,M_1\to M_2}=j_{R,M_1\to M_3}\;\;\forall\;\;
M_1<M_2<M_3
\ee
and thus we can form the inductive limit Hilbert space ${\cal H}^k_R$
which always exists, see appendix \ref{sb} and induces isometric injections 
$j_{R,M}:\; {\cal H}_{R,M}\to {\cal H}^k_R,\;M=(k,n)$ satisfying the 
consistency condition
\be \label{4.121a}
j_{R,M'}\cdot j_{R,M\to M'}=j_{R,M}\;\;\forall\;\;
M<M'
\ee
Equivalently we may consider the partially ordered and directed family of 
measures\\ 
$\nu_{R,M}=\langle \Omega_{R,M},\;.\;\Omega_{R,M}\rangle_{{\cal H}_{R,M}}$ 
which by construction are cylindrically consistent 
\ba \label{4.122}
\nu_{R,M'}(w_{R,M'}[I_{R,M\to M'} \cdot f_{R,M}])
&=& \langle \Omega_{R,M'},w_{R,M'}[I_{R,M\to M'}\cdot f_{R,M}]
\Omega_{R,M'}\rangle_{{\cal H}_{R,M'}}
\nonumber\\
&=& \langle j_{R,M\to M'}\Omega_{R,M},
j_{R,M\to M'} w_{R,M}[f_{R,M}]\Omega_{R,M}\rangle_{{\cal H}_{R,M'}}
\nonumber\\
&=& \langle \Omega_{R,M},
w_{R,M}[f_{R,M}]\Omega_{R,M}\rangle_{{\cal H}_{R,M}}
\nonumber\\
&=&\nu_{R,M}(w_{R,M}[f_{R,M}])
\ea  
for all $M<M'$.
Thus, by suitable extension theorems \cite{Yam75} a $\sigma-$additive
measure $\nu^k_R$ on the projective limit 
\begin{flalign} \label{4.123}
\gamma^k_R=\{(\phi_{R,M})_{M\in \mathbb{N}};&\;\phi_{R,M'}[I_{R,M\to M'}\cdot
f_{R,M}]=\phi_{R,M}[f_{R,M}] \;\nonumber \\
&\; \forall\; M=(k,n),\;M'=(k,n'),\;n<n',\;\; 
f_{R,M}\in L_{R,M}\}
\end{flalign}
exists and we have ${\cal H}^k_R\equiv L_2(\gamma^k_R,\; d\nu_R)$.
We call the fixed point {\it universal} if ${\cal H}_R={\cal H}^k_R\; \forall\;
k\in \mathbb{N}$ is independent of $k$ which is an expected 
property because we associate ${\cal H}_R^k$ with the 
continuum limit $\lim_{n\to \infty} {\cal H}_{R,k 2^n}$ and that limit 
is expected to be independent of $k$. In fact, an even higher degree of
universality is believed to occur: For instance, nothing is sacred about the 
factor of $2$ in our coarse graining map $I_{R,M\to 2M}$, we could have 
studied the flow for any $I_{R,M\to q M},\; q\in \mathbb{N},\; q>1$ and 
would expect independence of the inductive or projective limit of $q$.
In principle we can even do this for all $M<M',\; M,M'\in \mathbb{N}$ although
then the computation of the flow becomes much more technically involved.
These kinds of universality properties are studied in our companion paper
\cite{21}.

Likewise, we have a family of Hamiltonian operators satisfying the condition
\be \label{4.124}
j_{R,M\to M'}^\dagger\; H_{R,M'}\; j_{R,M\to M'}=H_{R,M}
\ee
for all $M<M'$. This reminds of the structure of a partially ordered and 
directed system of densely defined operators $A_M$ with domains $D_{R,M}$ 
such that $j_{R,M\to M'} D_{R,M}\subset D_{R,M'}$ for which one has 
the condition for all $M<M'$
\be \label{4.125}
A_{M'}\cdot j_{R,M\to M'}= j_{R,M\to M'}\cdot A_M
\ee
see appendix \ref{sb}. Such an inductive system of operators always has a densely defined
inductive limit $A$ on the corresponding inductive limit Hilbert space
satisfying $A j_{R,M}=j_{R,M} A_M$.
However, (\ref{4.125}) is much stronger than (\ref{4.124}): Multiplying 
(\ref{4.125}) from the left with $j_{R,M\to M'}^\dagger$ we indeed 
obtain (\ref{4.124}) due to isometry 
$j_{R,M\to M'}^\dagger j_{R,M\to M'}=1_{{\cal H}_{RM}}$
for $A_M=H_{R,M}$. However, multiplying (\ref{4.124}) from the left 
with $j_{R,M\to M'}$ only yields 
\be \label{4.126}
p_{R,M\to M'}\; H_{R,M\to M'} \; j_{R,M\to M'}
=j_{R,M\to M'} \; H_{R,M},\;\;p_{R,M\to M'}=j_{R,M\to M'} 
j_{R,M\to M'}^\dagger 
\ee
which yields (\ref{4.125}) iff $[p_{R,M\to M'},H_{R,M'}]=0$ since
due to isometry $p_{R,M\to M'} j_{R,M\to M'}=j_{R, M\to M'}$. 
Indeed, it would have been physically wrong if the renormalisation conditions 
would yield the structure of an inductive system of operators which always 
has an inductive limit $A$ because $A j_{R,M}=j_{R,M} A_M$ means that $A$
preserves all the subspaces $j_{R,M} {\cal H}_{R,M}$ of ${\cal H}_R$ separately
while we would generically expect correlations (non-vanishing 
matrix elements) between arbitrary pairs
of such subspaces. Moreover, we would have completely sidestepped Haag's
theorem \cite{Haag92} which states that different Hamiltonians generically 
cannot be defined on the same Hilbert space. Indeed, all that we are granted to 
obtain, if a fixed point of OS data exists,
is a quadratic form $H^k_R$ densely defined on the finite linear span 
of the vectors of the form $j_{R,M} \psi_{R,M},\; \psi_{R,M}\in 
D_{R,M}\subset{ \cal H}^k_{R,M}$ with $M=(k,n)$, denoted by $D^k_R$. 
On the other hand, it is 
maybe not completely hopeless that this quadratic form can be extended as 
a self-adjoint operator on ${\cal H}^k_R$ because $H^k_R$ is symmetric and 
positive. Namely, if it is in fact closable\footnote{That is, $D^k_R$ can 
be extended to a complete space with respect to the norm 
$||.||^k_R:=\sqrt{\langle .,H^k_R\;.\rangle_{{\cal H}^k_R}}$ and $H^k_R$ stays positive 
on that extension.} it has a unique Friedrichs extension 
as a self-adjoint positive operator \cite{RS80}. This hope is based on 
the fact that the renormalisation flow carefully keeps track of the vacua 
annihilated by the finite resolution Hamiltonians. We include 
it in our definition of universality if $H^k_R=H_R$ is independent of $k$
which is again plausible since we expect $H^k_R$ to be associated with
$\lim_{n\to \infty} H_{R,k 2^n}$.

\section{Outlook: Hamiltonian Renormalisation of Generally
Covariant Theories}
\label{s6}

In this paper we exploited the 1-1 correspondence between (Euclidian)
time translation invariant, time reflection invariant and reflection positive 
measures $\mu$ (OS measures) on the one hand and their corresponding 
OS data $({\cal H}, \Omega,H)$ in order to motivate the definition of a 
direct Hamiltonian flow of the OS data. This correspondence uses the OS
reconstruction and the Wiener measure construction algorithms respectively.
On the path integral side, renormalisation involves the introduction 
of families of finite resolution measures $M\mapsto \mu_{RM}$ where
$R$ is a fixed spatial IR cut-off and $M$ plays the role of an UV 
cut-off which labels the family. Upon picking a spatial block spin 
transformation also called a coarse graining map which preserves 
the space of OS measures, the path integral 
flow generates a sequence of measures $n\mapsto \mu^{(n)}_{RM}$ by defining
the measure with parameters $(n+1,M)$ from the measure with parameters 
$(n,2M)$. By OS construction we may define the OS data 
$({\cal H}^{(n)}_{RM},\; \Omega^{(n)}_{RM},\; H^{(n)}_{RM})$ from 
$\mu^{(n)}_{RM}$. We showed how the OS data with parameters $(n+1,M)$ and 
$(n,2M)$ are related. This relation, if one wants to stay completely 
within the Hamiltonian setting, is unsatisfactory in the sense that one 
needs to construct the Wiener measure for the OS data with parameters      
$(n,2M)$ in an intermediate step, a complicated step 
which one would like to avoid. However, that relation suggests a closely 
related modified flow that directly relates the OS data with parameters 
$(n+1,M)$ and $(n,2M)$ respectively without recourse to the Wiener measure. 
This direct flow thus gives directly a sequence 
$({\cal H}^{\prime(n)}_{RM},\;\Omega^{\prime(n)}_{RM},\; 
H^{\prime(n)}_{RM})$ of OS data from which one could construct a 
family of OS measures $\mu^{\prime(n)}_{RM}$ using the Wiener measure
construction. However, the flows are different, neither the primed and 
unprimed measures nor their OS data coincide at any finite resolution scale
$M$. Indeed, the path integral flow produces, if it exists, a fixed point 
family of 
measures $M\mapsto \mu^\ast_{RM}$ which qualify as the cylindrical projections,
corresponding to the resolutions $M$, of 
a continuum measure $\mu^\ast_R$. This continuum measure can be obtained by 
computing the continuum limit $\mu^\ast_R=\lim_{M\to \infty} \mu^\ast_{RM}$.
On the other hand, the Wiener measures $\mu^{\prime\ast}_{RM}$ corresponding
to a fixed point family of OS data 
$({\cal H}^{\prime\ast}_{RM},\;\Omega^{\prime\ast}_{RM},
\;H^{\prime\ast}_{RM})$ which is obtained via the direct Hamiltonian flow 
do not satisfy the cylindrical consistency condition. However, one can 
compute the corresponding continuum limit 
$\mu^{\prime\ast}_R=\lim_{M\to \infty} \mu^{\prime\ast}_R$. If 
in fact $\mu^\ast_R=\mu^{\prime\ast}_R$ then the continuum limit of the 
fixed point of OS data are the OS data of $\mu^\ast_R$ and thus the 
two different flows still produce the same continuum theory.

In our companion paper \cite{21a} we will demonstrate that the two flows indeed produce the 
same 
continuum theory for the case of a free, massive scalar field in one spatial
dimension. Ideally one would like to have necessary and/or sufficient 
criteria at one's disposal, which would guarantee such a coincidence in the general case.
These criteria may impose restrictions on the class of coarse graining maps,
discretisations, field content and Hamiltonians. However, even if such criteria
cannot be found or if there are examples for which the continuum limits of the 
fixed points of the flows do not coincide, still the direct Hamiltonian 
flow may produce a continuum Hamiltonian whose finite resolution 
matrix elements have much improved properties as 
compared the naive discretisations that started the direct renormalisation 
flow. Thus, the stage is prepared to attack more difficult models 
and eventually background independent quantum gravity in order to remove (or rather fix point)
the quantisation ambiguities of the quantum dynamics \cite{QSDI,QSDII,QSDV,Thi06,GT12,AQG,LT16,Winkler1,Winkler2,Winkler3,
SahThiWin,DL17_a,DL17_b,Thi98_Length}.
In that respect note that we were careful to state all the ingredients of the renormalisation
flow in a background independent way, even if the tests we performed are 
in the context of free scalar QFT in Minkowski space. \\
\\
\\
{\bf\large Acknowledgements}\\
Part of this work was financially supported by a grant from the 
Friedrich-Alexander University to the Emerging Fields Project ``Quantum
Geometry'' under its Emerging Fields Initiative. 
K. L. thanks the German National Merit Foundation for financial support.  
T.L. thanks the Heinrich-B\"oll Foundation for financial support.

\begin{appendix}

\section{Inductive Limits}
\label{sb}

Here we just list elementary definitions and implications on the topic 
of inductive limits of Hilbert spaces and operators.
\begin{Definition} \label{defb.1} ~\\
Let $I$ be an index set partially ordered and directed by $<$.\\
i.\\
A system of Hilbert spaces $\{{\cal H}_i\}_{i\in I}$ is called an inductive 
family iff for each $i<j$ there exist isometric injections 
\be \label{b.1}
J_{ij}:\;{\cal H}_i \to {\cal H}_j,\; J_{ii}={\rm id}_{{\cal H}_i}
\ee
that are subject to the compatibility condition for each $i<j<k$
\be \label{b.2}
J_{ik}=J_{jk} \;J_{ij}
\ee
ii.\\
A family of operators $\{A_i\}_{i\in I_c}$ with dense domains $D_i$ 
for $i\in I_c\subset I$ co-final in $I$ (i.e. for each $i\in I$ there 
exists $j\in I_c$ such that $i<j$) is said 
to be an inductive system provided that for each $i<j$ both in $I_c$
\be \label{b.3}
J_{ij} D_i \subset D_j,\;\;A_j \; J_{ij}=J_{ij}\; A_i
\ee
\end{Definition}
\begin{Lemma} \label{lab.1} ~\\
i.\\
An inductive system of Hilbert spaces has an inductive limit, i.e. there 
exists a Hilbert space ${\cal H}$ and isometric injections 
$J_i:\; {\cal H}_i\to {\cal H}$ for each $i\in I$ such that for all $i<j$
\be \label{b.4}
J_j \; J_{ij}=J_i
\ee
The inductive limit Hilbert space is unique up to a unitary map.\\
ii.\\
An inductive system of operators has an inductive limit, i.e. there
exists an operator $A$ densely defined on a domain $D\subset {\cal H}$ 
where ${\cal H}$ is the inductive limit Hilbert space such that for each 
$i\in I_c$
\be \label{b.5}
J_i \; A_i=A\; J_i
\ee
If moreover $A_i$ is essentially self-adjoint with core $D_i$ then
so is $A$ on $D$.
\end{Lemma}
Proof:\\
i.\\
We define vectors $\psi_i\in {\cal H}_i,\; \psi_j\in {\cal H}_j$ to 
be equivalent iff for some and therefore any\footnote{Suppose 
$k'>i,j$ then we find $\hat{k}>k,k'$ and have 
$0=J_{k\hat{k}}(J_{ik}\psi_i-J_{jk}\psi_j)=
J_{i\hat{k}}\psi_i-J_{j\hat{k}}\psi_j=
J_{k'\hat{k}}(J_{ik'}\psi_i-J_{jk'}\psi_j)$
whence by injectivity $J_{ik'}\psi_i=J_{jk'}\psi_j$.}
$i,j<k$ we have $J_{ik} \psi_i=J_{jk} \psi_j$.
We consider the equivalence classes $[\psi_i]$ and equip them  
with the inner product
\be \label{b.6}
\langle [\psi_i],[\psi_j]\rangle:=\langle J_{ik} \psi_i, J_{kj}\psi_j\rangle_{{\cal H}_k}
\ee
where $k$ is any $i,j<k$. This is independent of the representative because 
for any $i,j<k'$ we find $k,k'<\hat{k}$ and have by isometry and consistency
\begin{align} \label{b.7}
\langle J_{ik} \psi_i,J_{jk} \psi_j \rangle &=   
\langle J_{k\hat{k}}\;J_{ik} \psi_i,J_{k\hat{k}}J_{jk} \psi_j\rangle=   
\langle J_{i\hat{k}}\psi_i,J_{j\hat{k}} \psi_j\rangle = \nonumber\\
& =\langle J_{k'\hat{k}}\;J_{ik'} \psi_i,J_{k'\hat{k}}J_{jk'} \psi_j\rangle=   
\langle J_{ik'} \psi_i,J_{jk'} \psi_j\rangle
\end{align}
We extend the scalar product to the space of the formal finite 
linear combinations of equivalence classes by sesquilinearity and complete it 
to obtain the inductive limit Hilbert space ${\cal H}$. The required 
maps are 
\be \label{b.8}
J_i \psi_i:=[\psi_i]
\ee
and for any $i<j$ we have 
\be \label{b.9}
J_j J_{ij}\psi_i=[J_{ij}\psi_i]=[\psi_i]=J_i \psi_i
\ee
since the vectors $\psi_i,J_{ij} \psi_i$ are equivalent (choose $k=j>i,j$).
The $J_i$ are injections since $J_i \psi_i=[\psi_i]=0$ means
that $J_{ij} \psi_i=0$ for some $i<j$ hence $\psi_i=0$. They are isometric 
since (pick $k=j>i$).
\be \label{b.10}
\langle J_i \psi_i,J_i\psi'_i\rangle_{{\cal H}}=\langle [\psi_i],[\psi'_i]\rangle_{{\cal H}}=
\langle \psi_i,\psi'_i\rangle_{{\cal H}_i}
\ee
Finally suppose that two inductive limits $({\cal H},\{J_i\}_{i\in I})$ and 
$({\cal H}',\{J'_i\}_{i\in I})$ have been found. We define
\be \label{b.11}
U:\;{\cal H}\to {\cal H}';\; J_i\psi_i\mapsto J'_i \psi_i
\ee
and extend by linearity to the dense domain of the finite linear 
combinations of the $J_i \psi_i$. It has the inverse on its image
\be \label{b.12}
U^{-1}:\;{\cal H}'\to {\cal H};\; J'_i\psi_i\mapsto J_i \psi_i
\ee
and is isometric there (pick any $i,j<k$)
\be \label{b.13} 
\langle U J_i \psi_i, U J_j \psi_j\rangle_{{\cal H}'}=
\langle J'_i \psi_i, J'_j \psi_j\rangle_{{\cal H}'}=
\langle J'_k J_{ik} \psi_i, J'_k J_{jk}  \psi_j\rangle_{{\cal H}'}=
\langle J_{ik} \psi_i, J_{jk}  \psi_j\rangle_{{\cal H}_k}=
\langle J_i \psi_i, J_j \psi_j\rangle_{{\cal H}}
\ee
and can therefore be extended to a unitary operator to all of ${\cal H}$ by 
continuity.\\
ii.\\
We define $D$ to be the finite linear combinations of the vectors 
$J_i \psi_i,\; \psi_i\in D_i$ with $i\in I_c$. For $i\not\in I_c$
we find $i<j\in I_c$ such that $J_i\psi=J_j J_{ij} \psi_i$, so that 
it is enough to define $A$ on the $J_i \psi_i,\;i\in I_c$.   
As the $D_i$ are dense in ${\cal H}_i$,
$D$ is dense in ${\cal H}$. Then we {\it define} for $i\in I_c$
\be \label{b.14}
A J_i\psi_i:= J_i A_i \psi_i
\ee
and extend by linearity. 

This definition is consistent for suppose that 
$J_i \psi_i=J_j\psi_j$ then we find $i,j<k$ all in $I_c$ 
such that $J_{ik}\psi_i=
J_{jk}\psi_j$ and  
\be \label{b.15}
A(J_i \psi_i-J_j\psi_j)
=A(J_k J_{ik} \psi_i- J_k J_{jk} \psi_j)
=J_k A_k (J_{ik} \psi_i-J_{jk} \psi_j)=0 
\ee

Finally, by the basic criterion of essential self-adjointness, 
we know that $(A_j\pm i {\rm 1}_{{\cal H}_j})D_j$ is dense in 
${\cal H}_j$. It follows that for any $j\in I_c$
\be \label{b.16}
[A\pm i {\rm 1}_{{\cal H}}]J_j D_j=J_j[A_j\pm i {\rm 1}_{{\cal H}_j}] D_j
\ee
is dense in $J_j D_j$, hence $[A\pm i {\rm 1}_{{\cal H}}]D$ is dense 
in $\cal H$ and $A$ is essentially self-adjoint.

\section{Notes on Nelson-Symanzik Positivity}
\label{sc}

In the main text we claimed that (\ref{NelsonSymanzik}) implies (\ref{2.6}). We sketch 
the proof given in \cite{simon73,simonhoeghkrohn72} and adapt it to our notation.
For all $t\in\mathbb R$, let $Q_t = \Delta(\mathfrak B)$ and $Q = \times_{t\in\mathbb R} Q_t$ and let $\Sigma$ be the Baire sets of $Q$, i.e. the smallest $\sigma$-algebra, such that all compactly supported continuous functions are measurable.
We will construct $\mu$ as a probability measure on $(Q,\Sigma)$.
By the Gel'fand isomorphism, we have $\mathfrak B \simeq C(\Delta(\mathfrak B)) \simeq L^\infty(\Delta(\mathfrak B),\mathrm d\nu)$.
For $f\in L^2(\Delta(\mathfrak B),\mathrm d\nu)$, the positivity criterion guarantees that $e^{-\beta H}$ is positivity preserving, i.e. $f\geq 0$ implies $e^{-\beta H} f \geq 0$.
Since $e^{-\beta H}$ is a contraction on $L^2(\Delta(\mathfrak B),\mathrm d\nu)$, by the argument in \cite{simonhoeghkrohn72}, the preservation of positivity ensures that it is also a contraction on $L^1(\Delta(\mathfrak B),\mathrm d\nu)$.
By duality, we obtain a corresponding positivity preserving contraction on $L^\infty(\Delta(\mathfrak B),\mathrm d\nu)$, which defines a map $A_t$ on $C(\Delta(\mathfrak B))$ by isomorphism.
In order to proceed, we quote the following lemma from \cite{simon73}:
Let $Y_1,\ldots,Y_n$ be compact Hausdorff spaces and let $\mu_n$ be a Baire measure on $Y_n$.
Moreover, for $i=1,\ldots,n-1$, let $X_i$ be a bounded, positivity preserving map from $C(Y_i)$ to $C(Y_{i+1})$ such that $X_i 1 = 1$.
Then there exists a unique measure $\hat\nu$ on $Y_1\times\ldots\times Y_n$ such that for any $f_1,\ldots,f_n$ in $C(Y_1),\ldots,C(Y_n)$, we have
\begin{equation}
\int f_1(y_1)\ldots f_n(y_n)\,\mathrm d\hat\nu = \int f_n X_{n-1} (f_{n-1} X_{n-2}(f_{n-2}\ldots f_2 X_1 f_1)\ldots)\,\mathrm d\nu_n \text{.}
\end{equation}
Given $t_1 < \ldots < t_n$, we apply this lemma to $\mu_n = \nu$, $Y_i = Q_{t_i} = \Delta(\mathfrak B)$ and $X_i = A_{|t_{i+1}-t_i|}$ in order to obtain measures $\mu_{t_1,\ldots,t_n}$ on $Q_{t_1}\times\ldots\times Q_{t_n}$.
These measures can be used to define the measure $\mu$ on $(Q,\Sigma)$ by first defining it for functions $F$ in $C(Q)$ that depend only on finitely many $q_{t_i} \in Q_{t_i}$:
\begin{equation}
\int F\mathrm d\mu := \int F(q_{t_1},\ldots,q_{t_n})\,\mathrm d\mu_{t_1,\ldots,t_n}
\end{equation}
This constitutes a positive linear functional on the set $\bigcup_n \bigcup_{t_1,\ldots,t_n} C(Q_{t_1}\times\ldots\times Q_{t_n}$, which lies dense in $C(Q)$ by the Stone-Weierstrass theorem.
The functional thus extends to $C(Q)$ and by the Riesz-Markov theorem, we obtain a unique measure $\mu$ on $(Q,\Sigma)$.
By construction, the positivity preservation guarantees that $\mu$ is a positive measure and since $\mu(Q) = \int_{\Delta(\mathfrak B)} \mathrm d\nu = \left\langle\Omega,\Omega\right\rangle = 1$, it is a probability measure.

\end{appendix}

\end{document}